\documentclass[%
reprint,
superscriptaddress,
 amsmath,amssymb,
 aps,
 prx,
longbibliography,
floatfix,
]{revtex4-2}
\usepackage{amsmath}
\usepackage{braket}
\usepackage{geometry}
\usepackage{amssymb}
\usepackage{amsfonts}
\usepackage{mathtools}
\usepackage{appendix}
\usepackage{url}
\usepackage[utf8]{inputenc}
\usepackage{array}

\usepackage[dvipsnames]{xcolor}\usepackage[draft]{changes} \usepackage{color}   
\usepackage{hyperref}
\hypersetup{
    colorlinks=true, 
    linktoc=all,     
    linkcolor=blue,  
}

 \usepackage{graphicx} 

\usepackage{orcidlink}

\begin{document}
\preprint{APS/123-QED}
\title{Impact of Josephson junction array modes on fluxonium readout}

\author{Shraddha Singh\orcidlink{0000-0002-4921-1410}}\thanks{Corresponding author: shraddha.singh@yale.edu}
\affiliation{Department of Applied Physics and Physics, Yale University, New Haven, Connecticut 06511, USA}
\affiliation{Yale Quantum Institute, Yale University, New Haven, Connecticut 06511, USA}
\affiliation{AWS Center for Quantum Computing, Pasadena, California 91125, USA}
\author{Gil Refael\orcidlink{0009-0007-4566-8441}}
\affiliation{Institute for Quantum Information and Matter,
California Institute of Technology, Pasadena, California 91125, USA}
\affiliation{AWS Center for Quantum Computing, Pasadena, California 91125, USA}
\author{Aashish Clerk\orcidlink{0000-0001-7297-9068}}
\affiliation{Pritzker School of Molecular Engineering, University of Chicago, Chicago, Illinois 60637, USA}
\affiliation{AWS Center for Quantum Computing, Pasadena, California 91125, USA}
\author{Emma Rosenfeld\orcidlink{0000-0002-4153-4805}}\thanks{Present address: Google Research}
\affiliation{AWS Center for Quantum Computing, Pasadena, California 91125, USA}

\begin{abstract}

    Dispersive readout of superconducting qubits is often limited by readout-drive-induced transitions between qubit levels. While there is a growing understanding of such effects in transmon qubits, the case of highly nonlinear fluxonium qubits is more complex. We theoretically analyze measurement-induced state transitions (MIST) during the dispersive readout of a fluxonium qubit. We focus on a new mechanism: a simultaneous transition/excitation involving the qubit and an internal mode of the Josephson junction array in the fluxonium circuit. Using an adiabatic Floquet approach, we show that these new kinds of MIST processes can be relevant when using realistic circuit parameters and relatively low readout drive powers. They also contribute to excess qubit dephasing even after a measurement is complete. In addition to outlining basic mechanisms, we also investigate the dependence of such transitions on the circuit parameters. We find that with a judicious choice of frequency allocations or coupling strengths, these parasitic processes can most likely be avoided.
\end{abstract}

\maketitle
\section{Introduction}

The fluxonium superconducting qubit, based on a Josephson junction shunted by a capacitor and a large inductance, has emerged as a promising platform for quantum information. It exhibits long lifetimes~\cite{high_coherence_2019, somoroff_millisecond_2023, single_cooper_pair, earnest_realization_2018}, and both single~\cite{zhang_universal_2021} and two-qubit gates~\cite{ding_high-fidelity_2023, zhang_tunable_2024} have been demonstrated with high fidelity, with potential room for even further improvements~\cite{nesterov_cnot_2022, nesterov_proposal_2021, dogan_two-Fluxonium_2023, rosenfeld_designing_2024, nguyen_blueprint_2022}. The inductive shunt is a crucial part of the fluxonium circuit, with the most common realization being a Josephson junction array (JJA). In regimes where internal array modes are not excited, the JJA can act as a linear superinductance (see e.g.~\cite{masluk_microwave_2012, wang_achieving_2024}).       

In addition to coherence and the ability to do high-fidelity gates, the ability to make fast and efficient measurements is crucial for any qubit platform. Similar to other superconducting qubits, dispersive readout (using a driven readout cavity) has been the standard choice for fluxonium readout 
(see e.g.~\cite{{zhang_universal_2021}}).  While such measurement schemes should ideally be quantum non-demolition (QND)~\cite{blais2021circuit}, several experiments have reported non-QND backaction (either enhanced qubit error or transitions to non-computational states) during fluxonium readout~\cite{ding_high-fidelity_2023, gusenkova2021quantum,vooldriving2018,voolnon2014}.  Recent theoretical work on driven transmon qubits has provided insights into these so-called measurement-induced state transitions (MIST), showing that multi-photon transitions can lead to resonant excitation of the transmon to higher levels (see e.g.~\cite{shillito2022dynamics,xiao2023diagrammatic,khezri2023measurement,cohen2023reminiscence,dumas2024unified,sank2016measurement}).

\begin{figure}
    \centering
    \includegraphics[width=\linewidth]{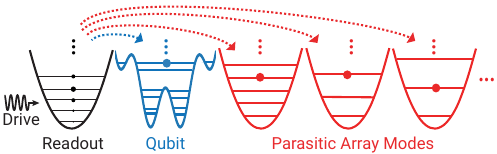}
    \caption{{\bf Schematic of a PMIST process.} A parasitic back-action effect where a readout drive applied to a cavity (black) simultaneously and resonantly excites extraneous linear modes (red) as well as the qubit (blue).
    }
    \label{fig:demo}
\end{figure}

Similar detrimental transitions can occur during fluxonium readout. The fluxonium circuit is fundamentally different and hence MIST effects here need independent analysis. For example, the enhanced nonlinearity can dramatically change the number and likelihood of potential transitions~\cite{nesterov2024measurement,xiao2023diagrammatic}. In this article, we analyze a unique MIST mechanism in fluxonium that arises due to the internal modes of the JJA during fluxonium readout (see Fig.~\ref{fig:demo}). We show that for realistic parameters and drive powers, deleterious resonant processes that simultaneously excite the qubit and an internal mode can occur. We term these processes parasitic MIST or PMIST. This work can be viewed as an example of a more general problem: how is the physics of MIST modified in the presence of a structured environment?

\begin{figure}[t]
\centering    
\includegraphics[width=\linewidth]{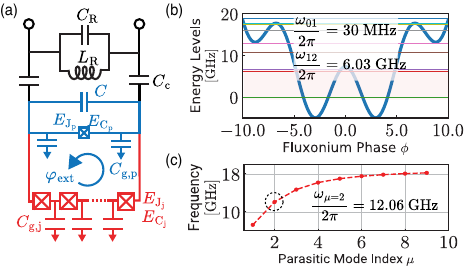}
\caption{{\bf Fluxonium readout circuit, qubit and array modes spectrum.} (a) The color scheme shows primary components that correspond to various modes, depicted in Fig.~\ref{fig:demo} when a JJA fluxonium circuit is connected to a readout cavity (R). The subscripts `$\mathrm{p,j}$' denote components of the phase-slip junction and the JJA, respectively. This circuit shows coupling capacitances ($C_\textrm{c}$), readout frequency parameters ($\omega_\textrm{r}=1/\sqrt{L_\textrm{R}C_\textrm{R}}$), parasitic ground capacitances in JJA ($C_\textrm{g,j}$) and next to the phase-slip junction ($C_\textrm{g,p}$). The differential capacitance $C$ adjusts the charging energy of the qubit mode (see Table~\ref{tab:readout_params}). (b) Fluxonium mode energy levels in units of $h$, with the highlighted area showing the first three levels essential for certain readout schemes~\cite{zhang_universal_2021}. Defining $\omega_{ij}$ as the splitting frequency between the fluxonium states $i,j$, we have $\omega_{01}/2\pi=30 \ \mathrm{MHz}, \ \omega_{12}/2\pi=6.03 \ \mathrm{GHz}$. (c) Parasitic mode frequencies $\omega_\mu/2\pi$. The lowest even mode $\mu = 2$ has the strongest coupling to the qubit (see Fig.~\ref{fig:coupling-strength} in App.~\ref{app:coupling}).
}
\label{fig:meas_circuit}
\end{figure}

We focus on a heavy fluxonium qubit operated at its flux sweet spot (cf~Fig.~\ref{fig:meas_circuit}) and investigate MIST, considering the coupling to the most relevant JJA internal mode. We treat the readout as an effective classical drive on the fluxonium-plus-JJA system and use an adiabatic Floquet branch analysis to identify dominant MIST processes. This method was used for transmon studies in Refs.~\cite{cohen2023reminiscence,dumas2024unified,khezri2023measurement}. We also validate this approach through full time-dependent simulations.  
Our work goes beyond showing that such processes could be relevant. We discuss how they provide a mechanism for degrading qubit coherence even after measurements are complete (via dephasing from dispersive couplings to the excited internal modes). We also discuss how alternate circuit designs affect PMIST processes, focusing on how modifications affect the parasitic mode to qubit coupling strengths. Our analysis suggests that in optimizing fluxonium readout, parasitic JJA modes introduce additional constraints on the circuit design. We stress that our analysis is not an exhaustive study of MIST effects in fluxonium readout but aims to show how new mechanisms involving internal JJA degrees of freedom arise in realistic setups.

The key contribution of this work is its identification of a novel, non-QND backaction on the qubit mode that differs fundamentally from previously studied MIST effects. Unlike prior analyses, which largely ignored the influence of the JJA parasitic modes or any additional non-qubit degrees of freedom, our study provides a detailed examination of these effects in the context of fluxonium qubits. While typically irrelevant for standard qubit operations, we show that the parasitic modes can induce a wide range of unexpected transitions and backaction mechanisms during readout, revealing a rich and underexplored source of error.

The remainder of this article is structured as follows. Sec.~\ref{sec:Fluxonium} analyzes the full circuit, including fluxonium, JJA, and readout cavity. Using a standard harmonic approximation~\cite{ferguson2013symmetries}, we derive qubit-parasitic mode coupling strengths and lower-bound parasitic mode effects during readout. Sec.~\ref{sec:MIST} analyzes readout dynamics, including MIST processes and dephasing from parasitic modes. Sec.~\ref{sec:expressions} examines the effects of varying coupling strengths between the qubit and a JJA internal mode on PMIST, and investigates different ranges of readout frequencies and parasitic mode frequencies using an energy-conservation picture. In the concluding Sec.~\ref{sec:conclusion}, we discuss directions for future work.

\renewcommand{\arraystretch}{1.5} 
\begin{table}[htb]
\centering
\begin{tabular}{|c|c|c|c|c|c|c|c|c|c|}
    \hline
     $N$ & $\varphi_{\textrm{ext}}$ & $E_{\textrm{J}_\textrm{p}}$ & $E_{\textrm{C}_\textrm{p}}$ & $E_{\textrm{C}}$ & $E_{\textrm{J}_\textrm{j}}$ & $E_{\textrm{C}_\textrm{j}}$ & $E_{\textrm{C}_\textrm{g,j}}$ & $E_{\textrm{C}_\textrm{g,p}}$ & $E_{\textrm{C}_\textrm{c}}$ \\
    \hline
    $122$ & $0.5\Phi_0$ & $7.30$ & $6.12$ & $1.73$ & $60$ & $0.74$ & $194$ & $1.94$ & $19.40$ \\
    \hline
\end{tabular}
\caption{{\bf Circuit parameters for Fig.~\ref{fig:meas_circuit}(a) inspired by Ref.~\cite{zhang_universal_2021}.} All energies are given in GHz. Here $\Phi_0=h/2e$ denotes the magnetic flux quantum. The capacitive energies $E_{\textrm{C}'}\mathrm{[GHz]}=\frac{e^2\times 10^{6}}{h}.\frac{1}{{C'}\mathrm{[fF]}}$ are computed from the corresponding capacitances $C'$ (see Table~\ref{tab:params} in App.~\ref{app:alt_circuits}). Here, $e,h$ are the fundamental constants, electron charge, and Planck's constant, respectively.}
\label{tab:circuit_params}
\end{table}

\renewcommand{\arraystretch}{1.5} 
\begin{table*}[tb]
    \centering
\begin{tabular}{|c|c|c|c|c|c|c|c|c|c|c|c|c|}
    \hline
    \textbf{Qubit ($\phi$) $\&$}&$\omega_{01}/2\pi$&$\omega_{12}/2\pi$ &$\tilde{E}^\phi_\textrm{C}$ &$g_{\phi \textrm{r}}/2\pi$&$\chi_{\phi \textrm{r}}(01)/2\pi$&$\chi_{\phi \textrm{r}}(12)/2\pi$&$\omega_\textrm{r}/2\pi$&$\kappa_\textrm{r}/2\pi$\\
    \cline{2-9}
\textbf{Readout ($r$)}&$30 \ \mathrm{MHz}$& $6.05 \ \mathrm{GHz}$ & $0.92 \ \mathrm{GHz}$& $25.50 \ \mathrm{MHz}$& $0.19 \ \mathrm{MHz}$&$0.10 \ \mathrm{MHz}$&$8.50 \ \mathrm{GHz}$&$1 \ \mathrm{MHz}$\\    
\hline\textbf{Parasitic-Mode} & \multicolumn{2}{c|}{} & $g_{\phi\mu}/2\pi$&$g_{\mu \textrm{r}}/2\pi$&$\chi_{\phi\mu}(01)/2\pi$&$\chi_{\phi\mu}(12)/2\pi$&$\omega_\mu/2\pi$&$Q_\mu$\\
    \cline{4-9}
\textbf{($\mu=2$)}&\multicolumn{2}{c|}{} &$315 \ \mathrm{MHz}$& $8.45 \ \mathrm{MHz}$& $-4.56 \ \mathrm{MHz}$& $20 \ \mathrm{MHz}$& $12.06 \ \mathrm{GHz}$&$10^{4}$\\    
\hline
\end{tabular}
\caption{{\bf Measurement parameters for qubit mode $\phi$, readout mode $\mathrm{r}$, and closest even parasitic mode $\mu=2$.} All quantities are derived and computed analytically using circuit parameters listed in Table~\ref{tab:circuit_params} (see App.~\ref{app:Hamiltonian} for details). \textbf{Qubit-Readout Parameters:} ($\omega_{if}$) qubit $i\rightarrow f$ splitting frequency between fluxonium excited states $i, f$; ($\tilde{E}^\phi_\textrm{C}$) qubit charging energy; ($g_{\phi \textrm{r}}$) qubit-readout coupling; ($\chi_{\phi \textrm{r}}(if)$) dispersive shift due to readout mode in the two-level $i-f$ system; ($\omega_\textrm{r}$) readout mode frequency; ($\kappa_\textrm{r}$) decay rate of the readout cavity. \textbf{Parasitic-Mode Parameters:} ($g_{\phi \mu}$) qubit-parasitic coupling; ($g_{\mu \textrm{r}}$) parasitic-readout coupling; ($\chi_{\phi \mu}(if)$) dispersive shift due to parasitic mode $\mu$ in the two-level $i-f$ system; ($\omega_\mu$) mode frequency; and ($Q_\mu$) internal quality factor inspired by~\cite{masluk_microwave_2012}. This internal quality factor is equivalent to a lifetime of $\sim 0.1\mu\textrm{s}$.}   \label{tab:readout_params}
\end{table*}

\section{Fluxonium Readout Circuit}\label{sec:Fluxonium}

We consider a JJA-fluxonium circuit dispersively coupled to a readout mode as shown in Fig.~\ref{fig:meas_circuit}. We choose circuit parameters (as listed in Table~\ref{tab:circuit_params}) motivated by recent experiments on heavy fluxonium~\cite{zhang_tunable_2024,zhang_universal_2021, ding_high-fidelity_2023}. We also restrict attention to the flux ``sweet spot" that maximizes qubit coherence. This choice is expected to reduce the number of allowed transitions in the circuit, as transitions between parity-conserving states via first-order processes are forbidden in this case~\cite{zhu_circuit_2013}. For our parameters, the qubit frequency ($\omega_{01}/2\pi$) is $\sim 30 \ \mathrm{MHz}$ and the plasmon frequency (i.e., splitting frequency between first and second qubit excited states $\omega_{12}/2\pi$) is $\sim 6 \ \mathrm{GHz}$ (see Table~\ref{tab:readout_params} for a full list of readout parameters).

Our work specifically investigates the role of the JJA, which comprises the inductive shunt of the fluxonium. The array comprises $N$ junctions and $N-1$ ground capacitances ($C_{\textrm{g}_n}$)~\cite{manucharyan2009fluxonium}. We neglect disorder effects~\footnote{The study on symmetries and disorders of the Josesphson junction array has been carried out in Ref.~\cite{ferguson2013symmetries}. More explicitly, the authors study the effect of a Guassian distribution of disorder along the array which essentially modifies the potential energy. The outcome of this disorder is a weak breaking of the symmetry resulting in coupling the qubit mode with all modes and not just even parasitic modes. Thus, a small disorder could lead to weak coupling between the qubit mode and the first parasitic mode which has the lowest frequency of $\sim 7$ GHz, and thus, can be more harmful for PMIST. However, we note that, for the parameters used in this work, calculations in Ref.~\cite{ferguson2013symmetries} show that the correction to the qubit-parasitic mode coupling due to disorder effects will be negligible.} and take junction parameters and parasitic ground capacitances to be uniform in the array (i.e., $C_{\textrm{g}_1}=..=C_{\textrm{g}_\textrm{N}}$). This capacitance value is given by $C_\textrm{g,j}$, where the subscript `$\mathrm{j}$' indicates the parasitic ground capacitance in the JJA~\footnote{Note that the ground capacitances $C_{\textrm{g}_n}$ are distinct from the self-capacitance $\frac{19.4}{E_{C,j}(\mathrm{GHz})}\mathrm{fF}$ (see Table~\ref{tab:params}) across the junctions in the array, which set the junction array plasmon frequency \cite{catelani2011relaxation}}. As shown in Fig.~\ref{fig:meas_circuit}(a), two additional identical ground capacitances $C_{\textrm{g}_0}, C_{\textrm{g}_\textrm{N}}$ near the phase-slip junctions in blue (see Fig.~\ref{fig:meas_circuit}) may have different values compared to those in the interior of the array, i.e., $C_{\textrm{g}_0}=C_{\textrm{g}_\textrm{N}}\equiv C_\textrm{g,p}\neq C_\textrm{g,j}$. Note that the subscript `$\mathrm{p}$' indicates the parasitic ground capacitances next to the phase-slip junction. The capacitance value $C_\textrm{g,p}$ can be adjusted by varying the distance between the capacitance pads during fabrication.

The JJA fluxonium circuit has $N$ internal degrees of freedom~\cite{ferguson2013symmetries}: one qubit mode ($\phi$) and $N-1$ internal modes ($\mu=1,2,...,N-1$). These internal modes are coupled via the ground capacitances and are referred to as the ``parasitic" modes of the JJA. 
In our notation, we label the readout mode as `$\mathrm{r}$'. The charge and flux quadratures of the qubit mode are denoted by $\hat N_\phi$ and $\hat \phi$ where $[\hat \phi,\hat N_\phi]=i\hbar$. 
We simplify the problem by treating all but the qubit mode as harmonic oscillators; this is a good approximation for standard device parameters~\cite{ferguson2013symmetries,viola2015collective,dumas2024unified}. We denote the photon loss and gain operators of the linear modes $\mathrm{r},\mu$ using $\hat a_\textrm{r},\hat a_\mu$ and $\hat a_\textrm{r}^\dagger,\hat a_\mu^\dagger$, respectively. 

Setting $\hbar=1$, the Hamiltonian of our fluxonium circuit has the form

\begin{equation}
   \hat H =\hat{H}_\phi + \hat{H}_\mu + \hat{H}_\textrm{r} + \hat{H}_{\textrm{int}},\label{Hamiltonian_total}
\end{equation}
where the qubit Hamiltonian $\hat{H}_\phi$ (with JJA inductive energy $E_\textrm{L}=E_{\textrm{J}_\textrm{j}}/N$) is 
\begin{equation}
\hat{H}_\phi / 2\pi = 4\tilde{E}^\phi_\textrm{C} \hat N_\phi^2+ E_{\textrm{J}_\textrm{p}}\cos{\hat\phi}+E_\textrm{L}\hat \phi^2 /2,\label{eq:Hphi}
\end{equation}
the junction array and readout Hamiltonians are $\hat{H}_\mu = \sum_{\mu}\omega_\mu \hat a_\mu^\dagger \hat a_\mu$ and $\hat{H}_\textrm{r} = \omega_\textrm{r} \hat a_\textrm{r}^\dagger \hat a_\textrm{r}$, respectively. Here, the second term in $\hat{H}_\phi$ has a `$+$' sign because we are operating at the sweet spot. The qubit charging energy
$\tilde{E}^\phi_\textrm{C}$ (see Table~\ref{tab:readout_params}) deviates from the target value of $E_\textrm{C}^{\phi}=\big(\frac{1}{E_\textrm{C}}+\frac{1}{E_{\textrm{C}_\textrm{p}}}+\frac{1}{2E_\textrm{g,p}}\big)^{-1}=1 \ \mathrm{GHz}$ due to parasitic capacitance. The coupling between the various modes is described by the interaction Hamiltonian
\begin{align}\label{eq:int_hamiltonian}
\hat{H}_{\textrm{int}} &= -i\sum_{\mu} g_{\phi\mu} \frac{\hat N_\phi}{{N_{\phi,\mathrm{ZPF}}}} (\hat a_\mu-\hat a_\mu^\dagger)\nonumber \\ &\quad-ig_{\phi \textrm{r}} \frac{\hat N_\phi}{{N_{\phi,\mathrm{ZPF}}}} (\hat a_\textrm{r}-\hat a_\textrm{r}^\dagger) \nonumber \\&\quad- \sum_{\mu} g_{\mu \textrm{r}} (\hat a_\textrm{r}-\hat a_\textrm{r}^\dagger)(\hat a_\mu-\hat a_\mu^\dagger).
\end{align}
For our parameters, the zero-point fluctuation value of qubit charge is $N_{\phi,\mathrm{ZPF}}=0.36$. Values for all remaining parameters are given in Table~\ref{tab:readout_params}. Here, the first, second and third terms represent the qubit-parasitic, qubit-readout, and parasitic-readout couplings, respectively. Explicit expressions for the $g_{\phi\mu}$ are discussed in Sec.~\ref{sec:expressions}.
 
We find that the lowest-frequency even parasitic mode $\mu=2$ has the strongest coupling to the qubit mode (see Fig.~\ref{fig:coupling-strength} in App.~\ref{app:coupling}); corresponding parameters are listed in Table~\ref{tab:readout_params}. The symmetry of the circuit in Fig.~\ref{fig:meas_circuit}(a) prevents coupling between all odd parasitic modes (including $\mu=1$) and other circuit modes~\cite{viola2015collective}. In App.~\ref{app:alt_circuits}, we derive the Lagrangians for two circuits, the symmetric circuit shown in Fig.~\ref{fig:meas_circuit}(a) and an alternate asymmetric circuit with a different grounding configuration, inspired by Ref.~\cite{zhang_universal_2021}. We show that the asymmetric circuit couples the qubit to the lowest frequency parasitic mode $\mu=1$, something that is likely even more detrimental for PMIST. Surprisingly, Fig.~\ref{fig:coupling-strength} in App.~\ref{app:coupling} shows that the $\mu=2,4,6$ parasitic modes couple to the qubit with a strength $g_{\phi\mu}$ that is stronger (larger in magnitude) than the qubit-readout coupling $g_{\phi \textrm{r}}$.

Given these insights, in the rest of this work, we focus on the symmetric circuit from Fig.~\ref{fig:meas_circuit} using parameters given by Tables~\ref{tab:circuit_params}-\ref{tab:readout_params} in Eq.~\ref{Hamiltonian_total}. Further, our description retains only the strongest coupled parasitic mode $\mu = 2$, along with the qubit and readout cavity. For details on other parasitic modes and their parameters, see App.~\ref{app:coupling}. Note that for our chosen parameters (see Table~\ref{tab:circuit_params}), the qubit couples roughly \emph{twelve} times more strongly to the parasitic mode at $\mu=2$ than it does to the readout $\mathrm{r}$~\footnote{In fact, the first four even parasitic modes have coupling strengths within a factor of $10$ of $g_{\phi \textrm{r}}$. See Fig.~\ref{fig:coupling-strength} in App.~\ref{app:coupling}}. This strong coupling implies that the parasitic mode can play a significant role in measurement-induced state transitions, i.e., the PMIST effect that is the subject of this work. In addition, coincidentally $\omega_{\mu=2}=2\omega_{12}$, but as we will show in this work, this is not an issue for the dominant PMIST processes when the parasitic mode is in the vacuum state. Nevertheless, once the parasitic mode is populated ($\braket{n_\mu}\ge 1$), due to a previous PMIST, it can impact the fluxonium qubit—introducing yet another pathway for MIST effects, which do not exist if JJA collective modes are neglected, in readout analysis.

\section{Parasitic-mode-Induced State Transitions: PMIST}\label{sec:MIST}

In this section, we analyze how the presence of a parasitic mode ($\mu=2$) affects the dynamics of a driven fluxonium circuit during a readout pulse. To simulate the linear drive on the readout cavity, we add a drive term $\hat{V}_\textrm{d}=-i\xi (\hat a_\textrm{r}-\hat a_\textrm{r}^\dagger)\cos{\omega_\textrm{d} t}$ to the system Hamiltonian in Eq.~\ref{Hamiltonian_total}. Considering the fluxonium qubit mode, parasitic modes, and readout cavity, a full numerical analysis of several excitations in the circuit would require a prohibitively large Hilbert space. To truncate our Hilbert space to feasible dimensions for numerical simulations, we include only a single parasitic mode $\mu=2$ (as previously justified in Sec.~\ref{sec:Fluxonium}) and replace the readout mode with a classical drive term~\cite{cohen2023reminiscence,dumas2024unified,khezri2023measurement} (see derivation in App.~\ref{app:semi-classical}). Under this semiclassical approximation, the driven circuit Hamiltonian includes the qubit mode $\phi$ and the parasitic mode at $\mu=2$, and is given by
\begin{align}
  \hat H_\textrm{s.c.}(\bar n_\textrm{r})=\hat H_0+\hat V_\textrm{s.c.}(\bar n_\textrm{r}).  \label{eq:drive_Ham}
\end{align}
Here, the bare Hamiltonian is
\begin{align}
\hat H_0=\hat H_\phi+\hat H_{\mu}+\frac{g_{\phi\mu}\hat N_\phi \hat N_\mu }{N_{\phi,\mathrm{ZPF}}N_{\mu,\mathrm{ZPF}}}, \label{eq:bare_ham} 
\end{align}
where we have used, $-iN_{\mu,\mathrm{ZPF}}(\hat a_\mu-\hat a_\mu^\dagger)=\hat N_\mu$, and the modified drive term $V_\textrm{s.c.}$ is
\begin{align}
    \hat V_\textrm{s.c.}(\bar n_\textrm{r})&=\frac{\xi_{\phi \textrm{r}}(\bar n_\textrm{r})}{N_{\phi,\mathrm{ZPF}}} \hat N_\phi\cos{\omega_\textrm{d} t}+\frac{\xi_{\mu \textrm{r}}(\bar n_\textrm{r})}{N_{\mu, \mathrm{ZPF}}} \hat N_\mu\cos{\omega_\textrm{d} t}\label{eq:drive},
\end{align}
where the effective drive amplitudes $\xi_{\mu(\phi) r}(\bar n_\textrm{r})=2g_{\mu(\phi) r}\sqrt{\bar n_\textrm{r}}$, and $\bar n_\textrm{r}$ denotes the average number of photons in the readout cavity. The zero-point fluctuation value of parasitic charge for $\mu=2$ is $N_{\mu,\mathrm{ZPF}}=1.58$. In the remaining text, we refer to the quantities $\xi_{\phi \textrm{r}/\mu \textrm{r}}$ as ``qubit drive strengths" and ``parasitic drive strengths", respectively.

Our primary focus is to analyze PMIST processes that introduce simultaneous transitions in the parasitic mode and the qubit mode. To identify the likely state transitions in the driven circuit, we first examine the energy eigenstates of the bare Hamiltonian $\hat{H}_0$ in Eq.~\ref{eq:bare_ham}. These states are hybridized fluxonium-parasitic mode states, and we label them as $\ket{\tilde{k}, \tilde{n}}$. A given state $\ket{\tilde{k}, \tilde{n}}$ corresponds to the eigenstate that has the maximum overlap with ``bare" fluxonium and parasitic mode states $\ket{k}_\phi\otimes\ket{n}_\mu$, i.e., the disjoint eigenstates of $\hat H_{\phi}+\hat H_{\mu}$. This state-labeling procedure corresponds to the \emph{overlap method}, which has been shown to be reliable for $n \ll 50$  in standard qubit-cavity systems~\cite{goto2025labeling}. In our work, the relevant transitions involve states with $n, k \ll 50$, and therefore do not face these limitations. We trust this method only for states whose overlap with the corresponding bare states is greater than $0.8$. However, our system differs from that of Ref.~\cite{goto2025labeling}, as the qubit is replaced by a multi-level fluxonium mode. Exploring the applicability of the alternative labeling strategies proposed in Ref.~\cite{goto2025labeling} to the joint fluxonium–parasitic mode Hilbert space would be a valuable direction for future work.

In what follows, we identify relevant state transitions in the driven fluxonium plus parasitic mode system. First, we perform an analysis based on the Floquet eigenstates of our system at a given fixed drive power~\cite{khezri2023measurement,cohen2023reminiscence,dumas2024unified}. We then simulate the drive ring-up to some chosen final photon number $\bar{n}_\textrm{r}$ and identify potential state transitions. We do this for a range of drive frequencies $\omega_\textrm{d}$. We find that the presence of the parasitic mode $\mu=2$ significantly increases the number of MIST processes in the system. We analyze the processes that cause these transitions and quantify their rates using perturbative approaches and Landau-Zener probability calculations~\cite{ikeda2022floquet}. We also show that the residual population in the parasitic modes, after a readout pulse, can lead to significant dephasing of the reset qubit mode, limiting the performance of the qubit for future use.

\subsection{Floquet branch analysis method} 
\begin{figure}[!htb]
    \centering
    \includegraphics[width=0.5\textwidth]{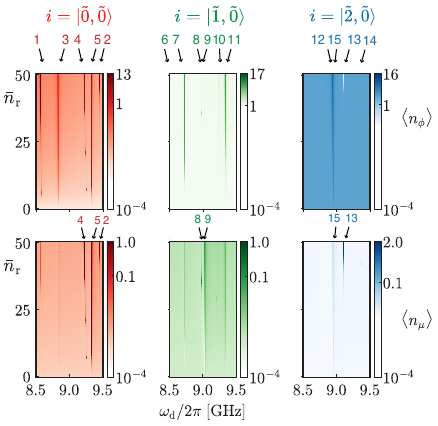}
    \caption{{\bf MIST and PMIST processes as seen in Floquet branch simulations.}  
    Each column corresponds to the branch associated with a specific undriven (but dressed) eigenstate $i=\ket{\tilde{k},\tilde{0}}$. The Floquet eigenstates were tracked with increase in $\bar n_r$ for this branch analysis where each step records the following quantities.
    \textbf{Top row:} Average fluxonium excitation number in the given branch $\langle n_\phi\rangle$, as a function of drive power ($\propto\bar{n}_\textrm{r}$) and drive frequency $\omega_\textrm{d}$. \textbf{Bottom row:} Average excitation number of the $\mu=2$ parasitic mode, $\langle n_\mu\rangle$. 
    Arrows and numbers indicate each transition (with numbers corresponding to Table~\ref{tab:PMIST}). The figures are plotted in logarithmic scale to pronounce the numbered streaks, the transitions of interest. Only sharp streaks indicate MIST or PMIST while any background change in color can be ignored. See Figs.~\ref{fig:Trans0}-\ref{fig:Trans2} of App.~\ref{app:Floquet-trans} for corresponding behavior of quasienergies.}
    \label{fig:Floquet}
\end{figure}
\begin{table*}[t]
    \centering
    \begin{tabular}{w{c}{2.0cm}w{c}{3.0cm}w{c}{2.0cm}w{c}{2.0cm}w{c}{2.5cm}w{c}{1.5cm}w{c}{2.5cm}}
\hline
\shortstack{\\\textbf{Transition }\\\textbf{No.}\\\textbf{ (see Fig.~\ref{fig:Floquet})}} &\shortstack{\\\textbf{Fluxonium}\\\textbf{MIST}\\\textbf{Process}} &\shortstack{\\\textbf{Threshold}\\\textbf{Drive}\\\textbf{Photon ($\bar n_\textrm{r}$)}}& \shortstack{\\\textbf{Drive}\\\textbf{Frequency}\\($\omega_\textrm{d}/2\pi$)}& \shortstack{\\\textbf{Quasienergy}\\\textbf{Gap}\\($\Delta_\textrm{ac}$)}& \textbf{PMIST}&\shortstack{\textbf{Drive Photons}\\\textbf{Absorbed}\\\textbf{(see Fig.~\ref{fig:trans_prof})}}\\
\hline
\rule{0pt}{4ex}$1.$&$\color{BrickRed}{\ket{\tilde{0},\tilde{0}}}$$\xleftrightarrow []{\hspace{1em}}\ket{\tilde{13},\tilde{0}}$&$13$ &$8.58 \ \mathrm{GHz}$&$0.96 \ \mathrm{MHz}$&$\times$ & $3$\\
$2.$&$\color{BrickRed}{\ket{\tilde{0},\tilde{0}}}$$\xleftrightarrow[]{\hspace{1em}}\ket{\tilde{13},\tilde{1}}$&$38$&$9.45 \ \mathrm{GHz}$&$0.35 \ \mathrm{MHz}$&$\color{BrickRed}{\checkmark}$ &$4$\\
$3.$&$\color{BrickRed}{\ket{\tilde{0},\tilde{0}}}$$\xleftrightarrow[]{\hspace{1em}}\ket{\tilde{8},\tilde{0}}$ &$\sim 0$&$8.83 \ \mathrm{GHz}$&$-$&$\times$ &$2$\\
$4.$&$\color{BrickRed}{\ket{\tilde{0},\tilde{0}}}$$\xleftrightarrow[]{\hspace{1em}}\ket{\tilde{6},\tilde{1}}^*$&$15$&$9.25 \ \mathrm{GHz}$&$0.50 \ \mathrm{MHz}$&$\color{BrickRed}{\checkmark}$ &$2$\\
$5.$&$\color{BrickRed}{\ket{\tilde{0},\tilde{0}}}$$\xleftrightarrow[]{\hspace{1em}}\ket{\tilde{3},\tilde{1}}^*$ &$6$&$9.33 \ \mathrm{GHz}$&$0.52 \ \mathrm{MHz}$&$\color{BrickRed}{\checkmark}$ &$2$\\
$6.$&$\color{ForestGreen}{\ket{\tilde{1},\tilde{0}}}$$\xleftrightarrow[]{\hspace{1em}}\ket{\tilde{17},\tilde{0}}$ &$32$&$8.51 \ \mathrm{GHz}$&$0.19 \ \mathrm{MHz}$&$\times$ & $4$\\
$7.$&$\color{ForestGreen}{\ket{\tilde{1},\tilde{0}}}$$\xleftrightarrow[]{\hspace{1em}}\ket{\tilde{7},\tilde{0}}$ &$5$&$8.73 \ \mathrm{GHz}$&$1.04 \ \mathrm{MHz}$&$\times$ & $2$\\
$8.$&$\color{ForestGreen}{\ket{\tilde{1},\tilde{0}}}$$\xleftrightarrow[]{\hspace{1em}}\ket{\tilde{12},\tilde{1}}$&$41$&$8.98 \ \mathrm{GHz}$&$0.53 \ \mathrm{MHz}$&$\color{ForestGreen}{\checkmark}$ & $4$\\
$9.$&$\color{ForestGreen}{\ket{\tilde{1},\tilde{0}}}$$\xleftrightarrow[]{\hspace{1em}}\ket{\tilde{2},\tilde{1}}$  &$2$&$9.03 \ \mathrm{GHz}$&$0.21 \ \mathrm{MHz}$&$\color{ForestGreen}{\checkmark}$ & $2$\\
$10.$&$\color{ForestGreen}{\ket{\tilde{1},\tilde{0}}}$$\xleftrightarrow[]{\hspace{1em}}\ket{\tilde{14},\tilde{0}}$ &$11$&$9.25 \ \mathrm{GHz}$&$0.74 \ \mathrm{MHz}$&$\times$ & $3$\\
$11.$&$\color{ForestGreen}{\ket{\tilde{1},\tilde{0}}}$$\xleftrightarrow[]{\hspace{1em}}\ket{\tilde{9},\tilde{0}}$ &$1$&$9.35 \ \mathrm{GHz}$&$0.61 \ \mathrm{MHz}$&$\times$ &$2$\\
$12.$&$\color{RoyalBlue}{\ket{\tilde{2},\tilde{0}}}$$\xleftrightarrow[]{\hspace{1em}} \ket{\tilde{12},\tilde{0}}$ &$2$&$8.95 \ \mathrm{GHz}$&$1.59 \ \mathrm{MHz}$&$\times$ & $2$\\
$13.$&$\color{RoyalBlue}{\ket{\tilde{2},\tilde{0}}}$$\xleftrightarrow[]{\hspace{1em}}\ket{\tilde{0},\tilde{2}}$&$14$&$9.10 \ \mathrm{GHz}$&$0.25 \ \mathrm{MHz}$&$\color{RoyalBlue}{\checkmark}$ & $2$\\
$14.$&$\color{RoyalBlue}{\ket{\tilde{2},\tilde{0}}}$$\xleftrightarrow[]{\hspace{1em}}\ket{\tilde{17},\tilde{0}}$&$11$&$9.35 \ \mathrm{GHz}$&$0.36\ \mathrm{MHz}$&$\times$ &$3$\\
$15.$&$\color{RoyalBlue}{\ket{\tilde{2},\tilde{0}}}$$\xleftrightarrow[]{\hspace{1em}} \ket{\tilde{10},\tilde{1}}$ &$23$&$8.98 \ \mathrm{GHz}$&$0.09 \ \mathrm{MHz}$&$\color{RoyalBlue}{\checkmark}$ & $3$\\
\hline
\end{tabular}
\caption{{\bf Measurement-induced-state-transition (MIST) observed in Fig.~\ref{fig:Floquet}.} Column $1$ lists the numbering used to mark the transitions in Fig.~\ref{fig:Floquet}. Here $\ket{\tilde{k},\tilde{n}}$ indicates the hybridized eigenstate of $\hat H_0$ (see Eq.~\ref{eq:bare_ham}) which has the maximum overlap with the state $\ket{k}_\phi\otimes \ket{n}_{\mu=2}$ in the disjoint Hilbert space of qubit mode ($\phi$) and parasitic mode ($\mu=2$). Column $2$ lists the MIST processes that start at the lowest average readout photon number $\bar n_\textrm{r}$ given by column $3$. These threshold values indicate the minimum drive amplitude that can cause the corresponding transition at the frequency $\omega_d$ given by the next column, and are predicted well using perturbative calculations as shown later in Fig.~\ref{fig:011} and Figs.~\ref{fig:Trans0}-\ref{fig:Trans2}.  In some cases, we use $\bar n_\textrm{r}\sim 0$ to indicate that the drive frequency is exactly resonant with the transition frequency between the two levels. A `$^*$'-marked state indicates hybridization at lower $\bar n_\textrm{r}$ due to preceding transitions~\footnote{$\ket{\tilde{6},\tilde{1}}\xleftrightarrow[]{\hspace{1em}}\ket{\tilde{3},\tilde{1}}$ at $\bar n_\textrm{r}\sim 0, \omega_\textrm{d}/2\pi\in[9.25,9.34] \ \mathrm{GHz}$.}. Column $4$ represents the drive frequency $\omega_\textrm{d}/2\pi$ at which these transitions occur. Column $5$ yields the quasienergy gap at the avoided crossing labeled as $\Delta_\textrm{ac}$. Column $6$ indicates if the process cannot occur without the parasitic mode, denoted as PMIST. The various colors for the checkmarks indicate whether the PMIST involves the state $\ket{\tilde{0},\tilde{0}}$ (red), $\ket{\tilde{1},\tilde{0}}$ (green) or $\ket{\tilde{2},\tilde{0}}$ (blue). Column $7$ indicates the number of drive photons ($\#$) involved in the energy-conserving process, illustrated in Fig.~\ref{fig:trans_prof}, which is responsible for these transitions. The title of the column \emph{drive photons absorbed} indicates the number of drive photons required for the transition in the joint Hilbert space of fluxonium and the single parasitic mode.}

    \label{tab:PMIST}
\end{table*}

Our first numerical analysis involves calculating the Floquet eigenstates of $\hat H_\textrm{s.c.}$ (see Eq.~\ref{eq:drive_Ham}) for various fixed values of the drive powers, as controlled by the average photon number $\bar n_\textrm{r}$. We do this by retaining the lowest 20 levels in the qubit subspace $\phi$ and 5 levels in the parasitic mode $\mu=2$~\footnote{We show that our results hold when simulated with 30 levels in the qubit mode and 10 levels in the parasitic mode. See Fig~\ref{fig:truncation} in App.~\ref{app:MIST}.}; the truncation for this analysis is discussed further in App.~\ref{app:numerics}. Our goal is to use these results to make predictions for a readout pulse involving a time-dependent drive power, identifying possible transitions starting from a dressed state $\ket{i} =\ket{\tilde{\phi}, \tilde{\mu}}$ where $\phi\in\{0,1,2\}$ and $\mu=0$ (i.e., the parasitic mode is initially empty). With $\omega_\mu/2\pi=12.06 \ \mathrm{GHz}$, the analysis in this section considers the regime of negative detuning where $\omega_{\mu=2}>\omega_\textrm{d}=\omega_\textrm{r} \gg \omega_\textrm{q}$, and can be replicated for any parasitic mode $\mu$. Note that we also analyze PMIST of an alternative circuit with $\omega_{\mu=2}/2\pi\sim 16 \ \mathrm{GHz}$ and $\omega_{01}/2\pi\sim 300 \ \mathrm{MHz}$ in App.~\ref{Will_circuit}.

We extract PMIST processes by tracking the evolution of the Floquet eigenstates as we increase the parameter $\bar{n}_\textrm{r}$, a method known as \emph{branch analysis}~\cite{dumas2024unified,cohen2023reminiscence}. We do this for a series of discrete values of $\bar{n}_\textrm{r}$ chosen to be integers. The simulation begins in a chosen eigenstate $\ket{i}_0$ of the bare Hamiltonian $\hat{H}_0 \equiv \hat{H}_\textrm{s.c.}[\bar{n}_\mathrm{r}=0]$ (see Eq.~\ref{eq:bare_ham}). Next, we compute the Floquet eigenstates $\{\ket{m}_{\bar n_\textrm{r}=1}\}$ of the Hamiltonian $\hat{H}_\textrm{s.c.}[\bar{n}_\textrm{r}=1]$, corresponding to a single photon increase in the readout cavity. We then identify the Floquet eigenstate of this Hamiltonian $\ket{i}_1$ that has maximum overlap with $\ket{i}_0$. We repeat this process iteratively, increasing $\bar{n}_\textrm{r}$ by one each time:
\begin{align}
\ket{i}_l:\max_{m}|\braket{i_{l-1}|{m}_{\bar n_\textrm{r}=l}}|^2.\label{eq:track_Floquet}   
\end{align}
We thus obtain a set of states $\ket{i}_0,\ket{i}_1,\ket{i}_2,...$ that we refer to as a branch. At a heuristic level, this trajectory of states describes the adiabatic evolution of the system as the drive power increases. The discrete steps in drive power that we consider in this trajectory emulate single-photon increases in the readout cavity photon number, i.e., $\delta \bar n_\textrm{r}=1$. Each step corresponds to an increase in drive power (see Eq.~\ref{eq:drive}) $\delta |\xi_{\mu (\phi),\textrm{r}}|^2=4g_{\mu (\phi),\textrm{r}}^2$. We make this choice to emulate the more quantum approach to branch analysis captured in Ref.~\cite{shillito2022dynamics,dumas2024unified}. 

The goal of the above procedure is to identify sudden changes in the number of excitations in the adiabatic Floquet eigenstates as drive power is increased. The use of a discrete step size also means that we are insensitive to transitions due to extremely narrow avoided crossings (something that would not be seen in an actual continuous power ramp at a finite rate). Note that in Eq.~\ref{eq:track_Floquet}, the overlaps between Floquet eigenstates are all computed at a fixed time within each drive period (i.e., at times $t_l = 2 \pi l/ \omega_\textrm{d}$)~\footnote{We have verified that our method can also reproduce the transmon-based results of Ref.~\cite{dumas2024unified} (where a time-averaged overlap was used).}.

For each state $\ket{i}_k$ in a given branch, we compute:
\begin{enumerate}
    \item The expectation value of the fluxonium excitation-number operator $\hat n_\phi=\sum_k k\ket{k}_\phi\bra{k}_\phi$, where $\ket{k}_\phi$ is the $k^\mathrm{th}$ {\it bare} fluxonium energy eigenstate,
    \item The expectation value of the parasitic-mode number operator $\hat n_\mu=\hat a_{\mu}^\dagger \hat a_{\mu}$ (for the single mode $\mu=2$ that we retain), and 
    \item The quasi-energy of the state $E_i \mod (\omega_\textrm{d}/2\pi)$.
\end{enumerate}
We can thus identify MIST and PMIST transitions by detecting sudden changes in the number of qubit or parasitic mode excitations as $\bar{n}_\textrm{r}$ increases, indicating an unwanted drive-induced hybridization of eigenstates.

\subsection{Branch analysis PMIST predictions}

Fig.~\ref{fig:Floquet} illustrates our main result, showing examples of PMIST drive-induced transitions, starting from initial (zero-drive) states that have maximum overlap with states $\ket{0}_\phi$, $\ket{1}_\phi$, and $\ket{2}_\phi$ in the fluxonium subspace and the ground state $\ket{0}_{\mu=2}$ in the parasitic subspace. For each branch, we use a color map to plot the average excitation number of the qubit mode $\langle n_\phi \rangle$ (top row) and the parasitic mode $\langle n_\mu \rangle$, for a range of readout drive frequencies (horizontal axes) and average photon numbers in the readout cavity $ \bar n_\textrm{r}$ (vertical axes). 

Results are shown for driven frequencies $\omega_\textrm{d} / 2 \pi$ in the range $8.5 - 9.5 \ \mathrm{GHz}$~\footnote{a relatively high-frequency choice, to reduce thermal, photon shot-noise induced dephasing in the qubit compared to lower-frequency bands}; other regimes are discussed in Sec.~\ref{sec:expressions}. For one-dimensional slices of the results at fixed $\omega_\textrm{d}$, along with the quasi-energies, see App.~\ref{app:Floquet-trans}. Any streak or sharp change in color intensity represents a sudden and significant jump in the qubit or parasitic mode population, i.e., MIST or PMIST. These transitions are identified as a sudden jump in the excitation value of the qubit or the parasitic mode between the observations at $\bar n_r-1$ and $\bar n_r$, where $\bar n_r$ is the threshold drive photon (see Table~\ref{tab:PMIST}). The parasitic transitions or PMIST correspond to simultaneous jumps in the population of the modes $\phi$ (Figs.~\ref{fig:Floquet}, top row) and $\mu=2$ (Figs.~\ref{fig:Floquet}, bottom row), as $\bar{n}_\textrm{r}$ varies. At these points, an avoided crossing in the quasi-energies of the Floquet states confirms the hybridization of the two states involved in the population exchange (see Figs.~\ref{fig:Trans0}-\ref{fig:Trans2} in App.~\ref{app:Floquet-trans}). Additional resonances may occur at alternate drive frequencies not shown in Fig.~\ref{fig:Floquet}. Table~\ref{tab:PMIST} lists significant transitions observed in our Floquet simulations and associated processes that cause them, identified through a perturbative analysis (see App.~\ref{app:stark-shift}) and energy conservation (shown later in Fig.~\ref{fig:trans_prof}). We also note that certain MIST processes, including PMIST, involve transitions at the flux sweet spot between parity-conserving states, due to virtual excitations via non-parity-conserving states.

The above results clearly show that coupling to the JJA parasitic modes enables new MIST processes beyond what would be predicted by a fluxonium-only simulation. A qualitatively new feature that arises due to the parasitic modes is the possibility of MIST-like transitions where the qubit loses excitations. For example, consider transition $\#13$ in Table \ref{tab:PMIST}. In this process, two drive photons are required, the fluxonium state has a downward transition $\ket{2}_\phi\rightarrow\ket{0}_\phi$, and the net energy released is used to excite the parasitic mode $\ket{0}_\mu\rightarrow \ket{2}_\mu$. By simple energy conservation, such an effect is not possible without a parasitic mode. PMIST processes like this can thus enhance errors within the qubit subspace, in addition to more standard processes which generate leakage. This error can be understood as an effective relaxation process where the qubit transitions from its first excited state to the ground state. See Fig.~\ref{fig:Trans2}, sub-panel ($13$) in App.~\ref{app:Floquet-trans}, for explicit population and quasienergy plots involving the two states.

Our results also display branch bunchings in addition to crossings in the negative detuning regime. This is qualitatively different from the transmon case studied in  Ref.~\cite{dumas2024unified}, where such branch bunching was only observed in the positive detuning $(\omega_\textrm{q}>\omega_\textrm{r})$ regime~\footnote{While less studied, branch bunching in the negative detuning can also occur in transmon systems but only at extremely high photon numbers (i.e. strong drives). This was learned through private referee communication}. Consider, for example, transitions $\#3$ and $\#4$ in Table \ref{tab:PMIST}. Here, we are driving at a frequency that exactly matches the transition frequency between two non-computational states, such that these states immediately start to hybridize into an equal superposition of the two 
original undriven states.  For example, in transition $\#4$, the readout drive frequency exactly matches the $\ket{3}_\phi \rightarrow \ket{6}_\phi$ transition frequency at zero readout excitations. In this case, levels $\ket{3}_\phi$ and $\ket{6}_\phi$ hybridize for any non-zero drive power $\bar n_\textrm{r}$. While this effect is not limited to PMIST, we highlight that the presence of parasitic modes can result in more exotic transitions. For example, in transition  $4$, one such transition between the states $\ket{\tilde{3},\tilde{1}}$ and $\ket{\tilde{6},\tilde{1}}$, where the parasitic mode was excited, eventually lead to a PMIST effect involving the computational state $\ket{\tilde{0},\tilde{0}}$. See Fig.~\ref{fig:Trans0}, sub-panel ($4$) in App.~\ref{app:Floquet-trans}, for explicit population and quasienergy plots involving the three states.

 \begin{figure}[t]
    \centering
    \includegraphics[width=\linewidth]{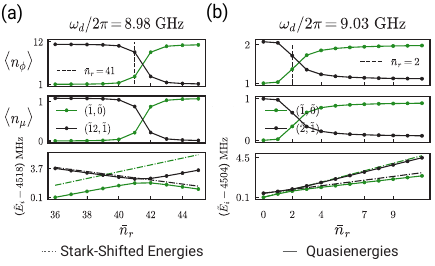}
    \caption{{\bf PMIST processes and eigenstate swapping.} Examples of PMIST corresponding to transitions \textbf{(a)} $8$ and \textbf{(b)} $9$ in Table~\ref{tab:PMIST}, relevant when starting from the undriven dressed state 
    $\ket{\tilde{1},\tilde{0}}$.
    \textbf{Top row:} Qubit mode average occupation $\braket{n_\phi}$
    in the two relevant Floquet adiabatic eigenstates, as a function of drive power $\bar{n}_r$.  \textbf{Middle row:} Parasitic mode average occupation $\braket{n_\mu}$. \textbf{Bottom row:} $\tilde{E}_i=E_i \ \textrm{mod} \ (\omega_\textrm{d}/2\pi)$ where $E_i$ is the Stark-shifted energy (dashed) obtained from first-order perturbation theory, or quasi-energy (solid) obtained from Floquet simulations showing avoided crossings. Plots are extracted from numerical data used in Fig.~\ref{fig:Floquet}. The data points are connected by lines for visual aid.}
    \label{fig:011}
\end{figure}

Our findings reveal that for our specific circuit choice, JJA parasitic modes can become significantly populated, as we show in Fig.~\ref{fig:Floquet}. For further insights, we now examine how the quasienergies and excitation numbers change as a function of drive power $\bar{n}_\textrm{r}$ when we pass through an avoided crossing associated with a PMIST process. Figs.~\ref{fig:011}(a,b) show how average qubit and parasitic mode excitation numbers change with $\bar n_\textrm{r}$ for fixed drive frequency corresponding to $8,9$ (respectively) in Table~\ref{tab:PMIST}. Both these transitions involve starting in the qubit's first excited state (i.e., branches associated with the undriven state $\ket{\tilde{1},\tilde{0}}$). The simultaneous exchange of population in the qubit mode $\phi$, shown in the top panels, and the parasitic mode $\mu=2$, shown in the middle panels, confirms that the transitions are indeed PMIST.

The bottom panel of Fig.~\ref{fig:011} shows the corresponding behavior of the branch quasi-energies as a function of drive power, for the same drive parameters. We plot both quasi-energies from the Floquet calculations and the predictions of a perturbative calculation including drive-induced Stark shifts (see App.~\ref{app:stark-shift} for details). For Fig.~\ref{fig:011}, the perturbative calculations predict the onset of an avoided crossing at a drive power close to the value derived from the Floquet simulations. Note that transition $\#9$ occurs at a very low average readout cavity photon number $\bar n_\textrm{r}=2$.

\subsection{Transition probability}\label{sec:LZ}
\begin{figure}[t]
    \centering
    \includegraphics[width=\linewidth]{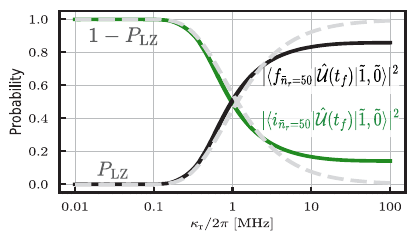}
    \caption{
        {\bf PMIST transition probabilities as a function of readout cavity ring-up rate.}
        We plot the probabilities for adiabatic (green) and diabatic (black) transitions for a time-dependent ring-up of the average cavity photon number from $\bar{n}_\textrm{r} = 0$ to $\bar{n}_\textrm{r} = 50$, for different choices of the cavity damping rate $\kappa_\textrm{r}$, which controls the speed of the sweep (see text). We start the system in the qubit's first excited state $\ket{\tilde{1},\tilde{0}}$. The drive frequency is $\omega_\textrm{d}/2\pi=9.05 \ \mathrm{GHz}$, corresponding to the crossing shown in Fig.~\ref{fig:011}(b). Here, $\ket{f_{\bar n_\textrm{r}=50}}$ and $\ket{i_{\bar n_\textrm{r}=50}}$ are the final states at the end of branch analyses for $\ket{\tilde{2},\tilde{1}}$ and $\ket{\tilde{1},\tilde{0}}$, respectively. The time-evolution operator is denoted by $\hat{\mathcal{U}}(t_\textrm{f})=\mathcal{T}\exp\big(-i\int^{t_\textrm{f}}_{0} \hat H_\textrm{s.c.}(t)dt\big)$, where $\mathcal{T}$ indicates time-ordering and $t_\textrm{f}=10/\kappa_\textrm{r}$. The adiabatic (green) curve corresponds to the probability of the occurrence of PMIST. We also plot the predictions of the Floquet branch analysis combined with a Landau-Zener approximation for the probabilities (gray), which are in excellent agreement over a wide range of $\kappa_\textrm{r}$.}
    \label{fig:LZ}
\end{figure}

Our Floquet branch analysis gives strong evidence that MIST and PMIST transitions may occur during the ring-up of the readout cavity during a readout pulse.  Here, we validate this approach by focusing on a specific transition, and we show (via explicit time-dependent simulations) that it occurs as predicted.  We also show that this full-time-domain simulation is in agreement with a Landau-Zener analysis that takes as its input the result of the Floquet branch analysis from the last subsection.  

We focus on the PMIST transition shown in Fig.~\ref{fig:011}(b), which corresponds to an avoided crossing energy gap of $\Delta_\textrm{ac}=0.21$ MHz. We perform a time-dependent simulation of $\hat H_\textrm{s.c.}$ in Eq.~\ref{eq:drive_Ham}, using drive powers determined by the time-dependent average readout cavity photon number in the form:
\begin{align}
    \bar n_\textrm{r}(t)&=\bar n_\textrm{r}(1-e^{-\kappa_\textrm{r} t/2})^2.\label{eq:LZ-n}
\end{align}
This corresponds to the ring-up~\footnote{Note that for the purpose of measurement, ring-up of the drive can be varied to avoid MIST effects, however, ring-down due to the decay of the readout resonator is the major concern causing MIST. We do not expect any difference in MIST effects emulated from ring-up and ring-down occurring at the same rate.} of a resonantly driven cavity with a damping rate $\kappa_\textrm{r}$. 

To calculate transition probabilities in this full time-dependent simulation, we initiate the system in the dressed state $\ket{\tilde{1},\tilde{0}}$, evolve under $\hat{H}_\textrm{s.c.}(t)$ from $t=0$ to $t=10 / \kappa_\textrm{r}$, and then compute the overlap of this state with the Floquet branches of the two states $\ket{i}=\ket{\tilde{1},\tilde{0}}$ and $\ket{f}=\ket{\tilde{2},\tilde{1}}$ associated with our predicted transition. We then repeat this calculation for different choices of $\kappa_\textrm{r}$, examining how the transition probabilities vary, with the results shown in Fig.~\ref{fig:LZ}. As expected, the probability of remaining adiabatic (green curve) decreases as one increases $\kappa_\textrm{r}$. Note that fully adiabatic evolution (i.e. the green curve) corresponds to a detrimental PMIST transition.  Adiabatic evolution means that starting in the bare state $\ket{\tilde{1},\tilde{0}}$, one follows $\ket{\tilde{1},\tilde{0}}$ branch as the drive amplitude is increased, ending up in $\ket{i}_{50}$ (the final state of this branch, see Eq.~\ref{eq:track_Floquet} for notation).  This state has a high overlap with the bare state $\ket{\tilde{2},\tilde{1}}$, indicating the qubit and internal mode have become excited.

For most values of $\kappa_\mathrm{r}$, the above probabilities are in agreement with the predictions of our Floquet branch analysis combined with a Landau-Zener approximation (see App.~\ref{app:LZ}) to the probability of a non-adiabatic transition~\cite{ikeda2022floquet,dumas2024unified}. These probabilities ($P_{\mathrm{LZ}}$) are shown in gray in Fig.~\ref{fig:LZ}. For large $\kappa_\textrm{r}$, we find that the probabilities do not saturate to $1$ and $0$ as would be expected in the standard Landau-Zener problem; they do, however, sum to unity (hence transitions to additional levels are not contributing). We attribute this small deviation to dressing effects that go beyond the standard Landau-Zener paradigm. Such discrepancies could often arise because (1) the effective time dependence of the Hamiltonian deviates from a simple linear magnetic field ramp, and because (2) transition probabilities are evaluated at finite evolution times—unlike the asymptotic limit \( t = -\infty \) to \( t = \infty \) assumed in the classic Landau-Zener formula~\cite{Vitanov,breuer1989adiabatic,BREUER1989507}.

\subsection{Post-readout qubit dephasing}\label{sec:dephasing}

The new PMIST processes we identify here can also potentially create errors \textit{after} the readout pulse is complete, as they lead to a new dephasing channel. A PMIST process results in a JJA parasitic mode having a residual excitation post-readout. As there is a non-zero dispersive coupling $\chi_{\phi\mu}$ between these modes and the qubit, and as these modes are believed to have relatively large internal quality factors $Q_\mu\sim 10^{4}$~\cite{masluk_microwave_2012, masluk2013reducing}, this will lead to the qubit acquiring a random phase (tied to the random time at which the parasitic mode relaxes). Below, we show master equation simulation results which quantify the scale of phase errors that would result from such processes.

\begin{figure}[t]
    \centering
    \includegraphics[width=\linewidth]{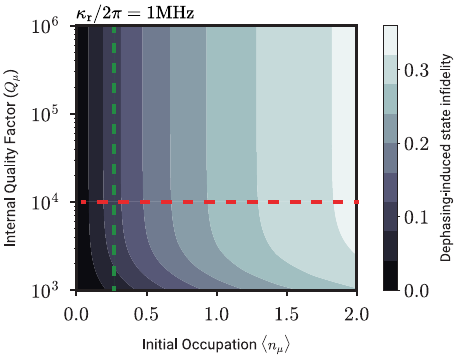}
    \caption{{\bf Contour plot for dephasing-induced state infidelity} due to random decay of an excited parasitic mode, after PMIST. The horizontal red line shows the parasitic internal quality factor quoted in~\cite{masluk_microwave_2012}. The green line shows the state infidelity due to dephasing for transition $\#9$ for various quality factors at $\kappa_\textrm{r}/2\pi=1 \ \mathrm{MHz}$ (see Figs.~\ref{fig:011}(b) and~\ref{fig:LZ}).}
    \label{fig:dephasing}
\end{figure}

We analyze how the decay of such a population dephases the fluxonium, by presenting an illustrative calculation where we initialize the parasitic mode in a thermal state~\footnote{We use a thermal state as the most general assumption on an imperfect parasitic mode ($Q_\mu\neq \infty$) after only the qubit mode has been reset. Correspondingly, the infidelities in the case of a coherent state will be increased by a factor of $2$.} with an average occupation number equal to $\braket{n_\mu}\in[0,2]$. This choice is based on Table~\ref{tab:PMIST}, showing that at a maximum drive power of $\bar n_\mathrm{r}=50$, parasitic mode populations can reach $\braket{n_\mu}$ = 2.0 after the readout pulse. We assume the qubit is rapidly prepared after readout in the state $\ket{+}=\frac{\ket{0}_\phi+\ket{1}_\phi}{\sqrt{2}}$. A master equation simulation illustrates the resulting qubit dephasing due to this mechanism. 

We investigate a reduced system of the parasitic mode $\mu=2$ and the fluxonium qubit (modeled as a two-level system), interacting under the dispersive Hamiltonian $\hat H_\theta/\hbar=\chi_{\phi\mu} \hat a_\mu^\dagger \hat a_\mu \sigma_\textrm{z}$, and with a loss dissipator having a collapse operator $\sqrt{\kappa_\mu}\hat a_\mu$, describing the parasitic mode internal loss. We let the system evolve from the above state for a time $T_\textrm{f}=10/\kappa_\mu$ long enough to allow the parasitic mode to relax, and then compute the fidelity of the final qubit state with the initial state $\ket{+}$, to quantify the errors induced by dephasing. This quantity is plotted in Fig.~\ref{fig:dephasing}, both as a function of the initial parasitic mode occupancy after readout $\braket{n_\mu}$ and its internal quality factor $Q_\mu$. 

The results of our simulations are shown in Fig.~\ref{fig:dephasing}. For concreteness, we consider a readout pulse with a frequency corresponding to the PMIST transition $\#9$ in Fig.~\ref{fig:011}(b) and a cavity damping rate $\kappa_\textrm{r}/2\pi=1 \ \mathrm{MHz}$ (which determines the ring-up time of the cavity to the maximum drive power). For these parameters, our previous simulations and Landau-Zener analysis suggest that at the end of the readout pulse, the parasitic mode will have an average non-zero excitation $\braket{n_\mu}=0.25$. Fig.~\ref{fig:dephasing} indicates that when using a realistic readout power, and for an internal quality factor $Q_\mu$ of $10^{4}$, the final post-readout parasitic mode population is sufficient to introduce phase errors with probability $\sim 0.1$. This indicates that PMIST processes could in principle contribute to a post-measurement loss of qubit coherence.  

\section{Effects of Circuit Modifications on PMIST}\label{sec:expressions}
Recognizing PMIST as a potentially significant qubit error channel, we now turn to mitigation strategies. Here, we discuss how adjusting the qubit frequency, readout cavity frequency and parasitic mode frequencies may affect the unwanted transitions. See App.~\ref{app:alt_circuits} for the Lagrangian and App.~\ref{app:coupling} for expressions quoted in this section. Note that due to the assumptions used in this work our analysis below is only valid for the junction count values $10^2<N<10^4$~\footnote{Note that the assumptions used in Ref.~\cite{viola2015collective} to derive the Hamiltonian were respected when analyzing the variation with $N$. The upper bound on $N\ll 8\pi^2 E_{\mathrm{C}_\mathrm{g,j}}\tilde{E}_\mathrm{C}^\phi\implies N\ll 16000$. Due to high nonlinearity of parasitic modes in the low $N$, and this restriction, our analysis is only valid for the region $10^2<N<10^4$.}.

\subsection{Root causes of PMIST}\label{sec:coupling}

We identify the primary causes of PMIST as,
\begin{itemize}
    \item high coupling-strength $g_{\phi\mu}$ (by analyzing the effect of setting different coupling strengths to zero in Floquet simulations), and
    \item low frequency-gap $\omega_\mathrm{r}-\omega_{\mu=2}$ (by enumerating MIST processes using energy conservation and verifying predictions against Table~\ref{tab:PMIST}).
\end{itemize}

\begin{figure}[t]
    \centering
    \includegraphics[width=\linewidth]{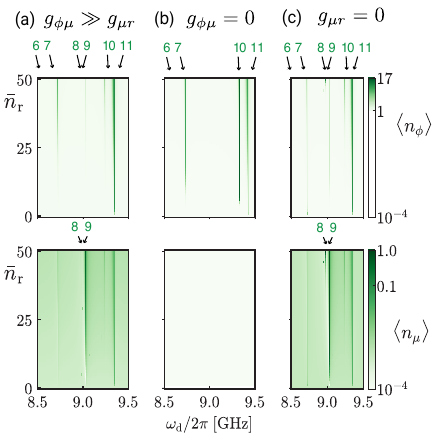}
    \caption{
        {\bf Sensitivity of PMIST processes to parasitic mode coupling strengths.} Panels show the result of Floquet branch analyses for the circuit parameters in Table~\ref{tab:circuit_params}. MIST processes observed in Fig.~\ref{fig:Floquet} are labeled with numbers and indicated with arrows. \textbf{(a)} All parameters are the same as in Fig.~\ref{fig:Floquet}(b). \textbf{(b)} Same, except now we set the parasitic mode to qubit coupling $g_{\phi \mu}$ to zero. Note that all PMIST features are now gone. \textbf{(c)} Same as (a), but we now set the parasitic mode to readout cavity coupling $g_{\mu \textrm{r}}$ to zero. Same as Fig.~\ref{fig:Floquet}, the figures are plotted in logarithmic scale to pronounce the numbered streaks, the transitions of interest. Only sharp streaks indicate MIST or PMIST while any background change in color can be ignored.} 
    \label{fig:coupling-Floquet}
\end{figure}

First, we compare the results of Floquet branch analyses for the initial state $\ket{\tilde{1}, \tilde{0}}$ under different coupling conditions, drive frequencies, and amplitudes. Fig.~\ref{fig:coupling-Floquet} identifies the main mechanism causing PMIST as the fluxonium-parasitic-mode coupling, $g_{\phi \mu}$. Fig.~\ref{fig:coupling-Floquet}(a) reproduces the Floquet simulation results from Fig.~\ref{fig:Floquet}. Using our previous choice of coupling strengths $g_{\phi\mu}$ and $g_{\mu \textrm{r}}$, PMISTs are shown as transitions $8$ and $9$ around $\omega_\textrm{d}/2\pi \sim 9$ GHz. In contrast, Fig.~\ref{fig:coupling-Floquet}(b) shows results from the same simulation, but with $g_{\phi \mu}$ set to zero. PMIST is absent, evident from the lack of parasitic transitions $(8,9)$ in the top panel and no streak or sharp change in color indicating parasitic mode excitations in the bottom panel. For $g_{\phi \mu}=0$, the parasitic mode population always remains below $\braket{n_\mu}=10^{-4}$. Therefore, we conclude that a finite $g_{\phi\mu}$ is the main mechanism behind PMIST: the readout cavity exchanges multiple excitations simultaneously with the parasitic mode and qubit, facilitated by the parasitic mode-qubit coupling $g_{\phi \mu}$. Further, Fig.~\ref{fig:coupling-Floquet}(c) shows that setting the parasitic-readout coupling to zero does not reduce PMIST. Thus, the qubit-readout coupling $g_{\phi \textrm{r}}$ alone does not cause significant transitions or PMIST processes without $g_{\phi \mu}$. Consequently, we turn to reducing the coupling strength $g_{\phi \mu}$ as the first strategy for reducing the likelihood of PMIST processes.

\begin{figure}[t]
    \centering
    \includegraphics[width=\linewidth]{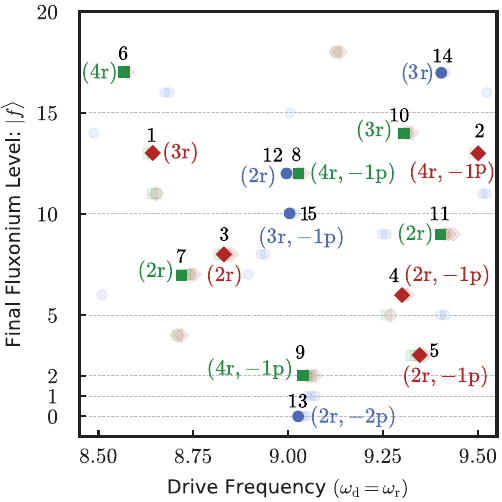}
    \caption{{\bf Energy-conserving processes $\ket{\tilde{i},\tilde{0}}\leftrightarrow\ket{\tilde{f},\tilde{y}}$ for Eq.~\ref{eq:En_cons} with $x\le 4, y\le 2, f\le 20$ and $i=0$ (red diamonds), $i=1$ (green squares), $i=2$ (blue circles).} The horizontal lines indicate the initial state $i$ for visual aid. Labels in black correspond to transition $\#$ listed in Table~\ref{tab:PMIST}. The parenthesized labels in the color of the form $(x \mathrm{r},- y\mathrm{p})$ denote the number of readout photons absorbed ($x$), from which energy equal to ($y$) parasitic mode photons are absorbed by the mode $\mu=2$, and the remaining energy is absorbed by the fluxonium mode $\phi$~\footnote{For example, transition $\#8$ corresponds to the absorption of four readout photons, which are converted into one parasitic-mode $\mu=1$ photon and an excitation from $\ket{0}_\phi$ to $\ket{12}_\phi$ in the fluxonium subspace.}. These labels correspond to the disjoint subspaces for simplified representation, but the energy conservation uses the eigenenergies of the hybridized eigenstates of $H_{0}$ (see Eq.~\ref{eq:bare_ham}). The faded points are weaker transitions not captured in the Floquet simulations.
}
    \label{fig:trans_prof}
\end{figure}

Next, we plot energy-conserving processes in Fig.~\ref{fig:trans_prof} that involve the parasitic mode $\mu=2$ to understand the spectrum of resonance conditions. We plot all processes that are approximately energy-conserving within a window $\epsilon$, i.e., that satisfy:
\begin{align}
    |x\omega_\textrm{r}-\tilde{\omega}_{if,y}|  \le \epsilon,
\label{eq:En_cons}
\end{align}
where $\tilde{\omega}_{if,y}$ is the transition frequency between the hybridized energy levels $\ket{\tilde{i},\tilde{m}}$ ($\ket{i}_\phi$) and $\ket{\tilde{f},\tilde{n}}$ such that $|m-n|=y$~\footnote{We can estimate resonance conditions by identifying energy-conserving processes, where $x$ drive photons are converted into a transition with an energy difference $\hbar\tilde{\omega}_{if,y}$ in the hybridized eigenspace of the fluxonium and parasitic mode $\mu=2$. This equation can also be interpreted as a process where $x$ readout photons convert into $y$ parasitic mode photons and a fluxonium excitation $\ket{i}_\phi \leftrightarrow \ket{f}_\phi$.}.
We assume a liberal value of $\epsilon = 25 \ \textrm{MHz}$ to identify processes that could become resonant once Stark shifts due to the readout drive, which are of order $\sim 25$ MHz (see Fig.~\ref{fig:stark-shift} in App.~\ref{app:stark-shift} for details), are accounted for. Fig.~\ref{fig:trans_prof} depicts all multi-photon processes that occur with approximate energy conservation, for $x \leq 4$, for drive frequencies within our target range, and when starting in one of the three lowest fluxonium levels~\footnote{Note that there are downward transitions from $\ket{2}_\phi$ (blue dots) to the states $\ket{1}_\phi$ (green line) and $\ket{0}_\phi$ (red line) in the fluxonium subspace in the presence of parasitic modes. One such example is captured by transition $\#13$ of Table~\ref{tab:PMIST}.}. We note that, in the presence of stronger parasitic-qubit coupling $g_{\phi\mu}$, the transitions shift very slightly towards lower drive frequency requirement. For example, in Fig.~\ref{fig:coupling-Floquet}(a), transition $\#6$ occurs at $\omega_d/2\pi=8.50 \ \mathrm{GHz}$ whereas in the absence of $g_{\phi\mu}$, in Fig.~\ref{fig:coupling-Floquet}(b), it occurs at $\omega_d/2\pi=8.56 \ \mathrm{GHz}$, as predicted by energy conservation in Fig.~\ref{fig:trans_prof}. 

Many of the transitions labeled in Fig.~\ref{fig:trans_prof} involving the parasitic mode have large transition numbers, requiring e.g. four readout photons. In general, for PMIST, several readout photons will be required to bridge an even larger energy gap between the readout frequency and parasitic mode frequency. In the perturbative regime, the larger the photon number in a transition, the lower its transition rate. As this gap increases, the likelihood of PMIST processes decreases~\cite{kurilovich2025high}. Thus, as the second strategy towards reducing PMIST, we discuss the effect of increasing the frequency gap between the readout mode and the lowest-frequency parasitic mode ($\omega_\textrm{r}-\omega_{\mu=2}$).

\subsection{Reducing the coupling strength $g_{\phi \mu}$}

Here, we analyze the dependence of these coupling strengths on circuit and readout parameters. As discussed, only even-index parasitic modes have a non-zero coupling to the qubit (see the Lagrangian derivation in App.~\ref{app:alt_circuits}). The coupling strength $g_{\phi \mu}$ between an even parasitic mode and the qubit follows~\cite{viola2015collective}, 
\begin{align}
\frac{g_{\phi\mu}}{2\pi}&=\frac{4}{\sqrt{2N}}\frac{\tilde{E}^\phi_\textrm{C}E^\textrm{e}_{\textrm{c},\mu}c_\mu}{E_{\textrm{C}_\textrm{g,j}}s_\mu^2} \cdot {N_\phi}_{\mathrm{ZPF}} \cdot {N_\mu}_{\mathrm{ZPF}},\label{eq:coupling}
\end{align}
where $c_\mu=\cos{\frac{\pi\mu}{2N}}, s_\mu = \sin \frac{\pi \mu}{2N}$. $\tilde{E}_\textrm{C}^\phi$ and $E^\textrm{e}_{\textrm{c},\mu}$ are the qubit and even parasitic mode charging energies, respectively, and $N_{\phi/\mu,\mathrm{ZPF}}$ are the zero-point fluctuation values for the qubit and parasitic modes.  $\tilde{E}_\textrm{C}^\phi,E_{\textrm{c},\mu}^\textrm{e}$ and $N_{\phi/\mu,\mathrm{ZPF}}$ are given in App.~\ref{app:coupling}, where
\begin{align}
\frac{1}{E_{\textrm{c},\mu}^\textrm{e}}&\propto\Big[\frac{1}{E_{\textrm{C}_\textrm{j}}}+\frac{1}{4E_{\textrm{C}_\textrm{g,j}}s_\mu^2}\Big].\label{eq:parasitic}
\end{align}
All the other variables represent independent quantities listed in Table~\ref{tab:circuit_params}.  

We see that suppressing the parasitic capacitance to ground near the junction array suppresses the qubit-parasitic coupling $g_{\phi\mu}$. However, this is constrained by practical limitations to order $\mathcal{O}(0.1) \ \mathrm{fF}$ per junction. The parasitic modes with the strongest coupling to the qubit have $\mu\ll N$. The large $N$, small $\mu$ limit with $c_\mu\approx 1$ yields
\begin{align}
    E^\textrm{e}_{\textrm{c},\mu}\approx 4E_{\textrm{C}_\textrm{g,j}}s_\mu^2, \quad \tilde{E}^\phi_\textrm{C}\propto \frac{1}{N^2}\implies g_{\phi\mu}\propto \frac{1}{N^{5/2}}.\label{eq:dep1}
\end{align}
These dependencies are plotted in Fig.~\ref{fig:circuit_comp} of App.~\ref{app:coupling}. We find that the coupling strength \textit{decreases} with the number of junctions $N$; however, a limit to this increase may be set by the requirement of a constant inductance $E_\textrm{L}=E_{\textrm{J}_\textrm{j}}/N$. Increasing \( N \) while keeping \( E_\textrm{L} \) and \( E_{\textrm{C}_\textrm{g,j}} \) constant also lowers the parasitic charging energy (see App.~\ref{app:coupling} for details), an undesirable consequence.
\subsection{Increasing the readout and parasitic frequency gap}\label{mode-frequencies}

We discuss tailoring the circuit to avoid resonance conditions required for PMIST.

\paragraph{Lowest parasitic-mode frequency ($\omega_{\mu=2}$):}
To mitigate PMIST, adjust $\omega_\mu$ so that $\omega_\mu \gg \omega_\textrm{r}$ for $\mu=2$, requiring more readout photons for PMIST and reducing transition rates. The parasitic mode frequencies for even $\mu$ are:
\begin{align}
    \frac{\omega_\textrm{e}}{2\pi}=\sqrt{8E_{\textrm{c},\mu}^\textrm{e} E_{\textrm{J}_\textrm{j}}}
\end{align}
See Eq.~\ref{eq:parasitic} for $E_{\textrm{c},\mu}^\textrm{e}$ and Table~\ref{tab:circuit_params} for $E_{\textrm{J}_\textrm{j}}$. In the large $N$ and small $\mu$ limit, the parasitic frequency decreases with increased junction count and parasitic ground capacitance. Decreasing parasitic ground capacitance (and thus increasing $E_{\mathrm{C}_\mathrm{g,j}}$), increases the parasitic frequency and decreases the qubit-parasitic coupling, desirably. Increasing $N$ reduces the coupling strength but decreases the gap between $\omega_\textrm{d}$ and $\omega_\mu$. Decreasing $N$ to increase this gap requires considering nonlinear corrections~\cite{viola2015collective} and fixed inductance, making the analysis complex.

\begin{figure}[t]
    \centering
    \includegraphics[width=\linewidth]{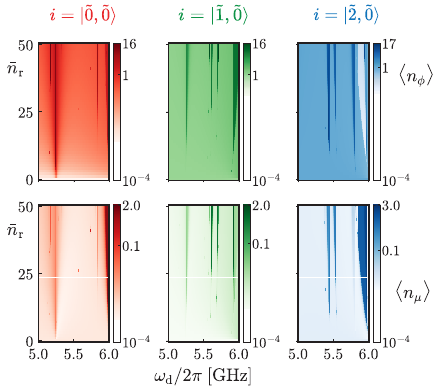}
    \caption{{\bf Floquet simulations at lower readout frequencies.} Circuit parameters are the same as in Tables~\ref{tab:circuit_params} and~\ref{tab:readout_params} for branch analysis starting in the dressed hybridized eigenstate $i=\ket{\tilde{k},\tilde{0}}$, with maximum overlap to the un-hybridized states $\ket{k}_\phi\otimes\ket{0}_{\mu=2}$. Same as Fig.~\ref{fig:Floquet}, the figures are plotted in logarithmic scale to pronounce the numbered streaks, the transitions of interest. Only sharp streaks indicate MIST or PMIST while any background change in color can be ignored.}
    \label{fig:Flo_low}
\end{figure}

\paragraph{Drive frequency ($\omega_\textrm{d}$):}
Reducing the readout cavity frequency increases the gap between readout and parasitic mode excitations, requiring more readout photons for PMIST and lowering transition rates. However, Floquet simulations in the low-frequency readout regime (Fig.~\ref{fig:Flo_low}) show higher parasitic mode populations compared to Fig.~\ref{fig:Floquet}. This is due to increased resonance density when $\omega_\textrm{d}/2\pi \approx 6$ GHz, matching the plasmon transition $\omega_{12}/2\pi$ of the fluxonium qubit mode and half the parasitic mode frequency. Reduced readout frequencies introduce multiple frequency collisions, increasing PMIST resonances. While these transitions are higher order (based on energy conservation), they still show up prominently in the branch analysis, suggesting they could be problematic. In addition to Floquet analysis, formal transition probability calculations, same as Sec.~\ref{sec:LZ}, are needed to predict the impact of these additional resonances. 

\section{Conclusion and Further Work}\label{sec:conclusion}
In this work, we have analyzed the impact of internal array degrees of freedom on driven JJA fluxonium qubits, showing that new pathways for measurement-induced state transitions can occur via excitations of JJA parasitic modes. These processes, which we have called PMIST, occur at particular resonance conditions when the energy of several readout photons is equal to a small number of parasitic mode excitations and a fluxonium mode excitation.

We find that PMIST transitions can occur at meaningfully high rates (with avoided crossing quasi-energy gap $\Delta_{\textrm{ac}}\sim 0.50 \ \textrm{MHz}$) because of the strong coupling between parasitic modes and the qubit. Such state transitions can occur even at low readout drive powers (corresponding to small intracavity photon numbers) while using typical parameters that enable high-fidelity, dispersive readout. For example, we have shown that PMIST does lower the onset of MIST processes to $\sim 2$ readout photons at certain drive frequencies. In addition, PMIST has the potential to significantly dephase the qubit post-measurement. This could in turn limit the qubit gate fidelities required for quantum error correction and ultimately the performance of a quantum processor. However, despite the strong parasitic mode and qubit coupling, we find that PMIST is unlikely to occur for the vast majority of readout cavity frequencies. Therefore, these processes can be avoided via a judicious choice of readout, junction array, and fluxonium parameters. We analyze the trend in PMIST for various drive frequencies, parasitic mode frequencies, coupling constants, grounding options and circuits with two different qubit frequencies equal to $\sim 30$ and $\sim 300 \ \mathrm{MHz}$.

We have presented a first analysis toward understanding the role of parasitic modes in the dispersive readout dynamics of a fluxonium circuit. Mitigating the parasitic mode excitations could involve careful selection of the readout cavity frequency and varying junction energies along the array to localize parasitic modes. While we have not explored this direction in our work, it is an intriguing future prospect. Such modifications alter the parasitic mode spectrum towards reducing excitation probability. The circuit parameters used in this work correspond to a parasitic mode of the fluxonium's junction array. However, other modes with similar frequencies or stronger coupling to the qubit may also participate in the environment of the fluxonium. The simulations with one parasitic mode at a time can be combined to get a better understanding of the landscape of spurious transitions involving the qubit, readout and any (one) parasitic mode. This may include, for example, confined package modes, slot line modes, and harmonics of coplanar waveguide resonators for readout. Our results show the significance of considering all such modes when driving many excitations into highly nonlinear circuits.
 \section{Code and Data Availability}
 Some of the simulations in this work use scqubits~\cite{groszkowski_scqubits_2021} and QuTip~\cite{johansson_qutip_2012} packages. The code and data are available via GitHub at Ref.~\cite{singh_shraggyreadout_simulations_2025}.

\section{Acknowledgments}
 We thank Akshay Koottandavida, Daniel K. Weiss, Sumeru Hazra, Connor Hann, Kyungjoo Noh and Simon Leiu for fruitful discussions. G.R. is grateful for support from the Simons Foundation as well as support from the IQIM, an NSF Physics Frontiers Center. We acknowledge the AWS quantum computing program where the initial steps of this project were envisioned. We thank Mostafa Khezri for feedback on the manuscript. 
\appendix
\section{Role of symmetry in readout circuits}\label{app:alt_circuits}
\begin{figure}[htb]
    \centering
    \includegraphics[width=\linewidth]{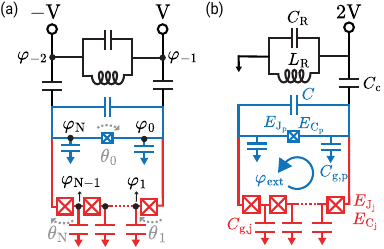}
    \caption{{\bf Alternative readout circuits.} (a) Symmetric circuit same as Fig.~\ref{fig:meas_circuit}(a), (b) Asymmetric circuit inspired by Ref.~\cite{zhang_universal_2021}. Alternative (b) requires a single-point connection to the readout line, unlike the symmetric circuit in (a). The symmetric circuit is used as a reference for denoting phase variables $\varphi_i$ at each node and phase difference variables $\theta_j$ across each junction, used in the derivation of the Lagrangian for both circuits.}
    \label{fig:circuit_choice}
\end{figure}
\begin{table}[htb]
    \begin{center}
    \begin{tabular}{|c |c| c |c| }
     \hline
     \textbf{Parameters} & \textbf{Variables} & \textbf{Values}\\ 
    \hline
    Phase-slip junction capacitance &$C_\textrm{p}$ &$3.17$ fF\\ 
    \hline
    Differential capacitance &$C$ &$11.27$ fF\\ 
    \hline
    JJA junction capacitance &$C_\textrm{j}$&$26.2$ fF\\ 
    \hline
    Phase-slip ground capacitance&$C_\textrm{g,p}$&$10 \ \mathrm{fF}$\\ 
    \hline
    JJA ground capacitance&$C_\textrm{g,j}$&$0.1 \ \mathrm{fF}$\\ 
    \hline
    Coupling capacitance&$C_\textrm{c}$ &$1 \ \mathrm{fF}$\\ 
     \hline
    Capacitance of resonator&$C_\mathrm{R}$&$294 \  \mathrm{fF}$\\
     \hline
      ZPF of resonator charge operator&$N_{\mathrm{r},\mathrm{ZPF}}$&$2.84$\\
     \hline
      ZPF of fluxonium charge operator&$N_{\phi,\mathrm{ZPF}}$&$0.36$\\
     \hline
      ZPF of parasitic charge operator&$N_{\mu=2,\mathrm{ZPF}}$&$1.58$\\
     \hline
    \end{tabular}
    \end{center}
    
    \caption{{\bf Capacitances and zero-point fluctuation (ZPF) values.} These capacitance values for the symmetric circuits are used throughout this article except App.~\ref{Will_circuit}. Here, $V_{\mathrm{ZPF}}=4E_{C_\mathrm{R}}N_{\mathrm{r},\mathrm{ZPF}}$ is used to obtain the qubit and parasitic drive strengths in Eq.~\ref{eq:drive_Ham}. For capacitive, inductive, and Josephson energy values, see Table~\ref{tab:circuit_params}.}
    \label{tab:params}
    \end{table}

The fluxonium readout circuit in Fig.~\ref{fig:meas_circuit} can be modified in several ways, affecting performance metrics. This circuit has a configuration with symmetry (identified in Ref.~\cite{ferguson2013symmetries}) that removes coupling to the lowest frequency mode $\mu=1$~\cite{viola2015collective}. Crucial to this symmetry are the equal coupling capacitances on both ends of the circuit (see Fig.~\ref{fig:circuit_choice}(a)). We present a modification with a different grounding option that does not respect this symmetry (see Fig.~\ref{fig:circuit_choice}(b)). This asymmetric circuit, inspired by Ref.~\cite{zhang_universal_2021}, is relevant for a hanger geometry with a single readout feedline. In this appendix, we derive the Lagrangian for both circuits and show that not preserving this symmetry can be detrimental to readout. Therefore, we use the symmetric circuit for fluxonium readout analyses in the rest of the appendix and the main text.
\subsection{Lagrangian}

To derive the Lagrangians, we follow the recipe in Refs.~\cite{viola2015collective,ferguson2013symmetries} for the symmetric circuit and adapt it for the asymmetric circuit in Fig.~\ref{fig:circuit_choice}(b). Importantly, we show that the symmetry in the circuit~\ref{fig:circuit_choice}(a), which prevents any coupling with the lowest-frequency parasitic mode ($\mu=1$), is not preserved in~\ref{fig:circuit_choice}(b). The presence of coupling to this mode is detrimental to readout because of the low-frequency gap between the readout and the parasitic mode, as discussed in Sec.~\ref{sec:expressions}.

The Lagrangian corresponding to both circuits is a combination of the Lagrangians, $\mathcal{L}_\textrm{PS}$ from the phase-slip junction shown in blue (comprising of the junction with $E_{\textrm{J}}/E_{\textrm{C}}\sim 1$ and the differential capacitor $C$), $\mathcal{L}_\textrm{JJA}$ from the (red) junctions in the array, $\mathcal{L}_\textrm{g}$ from the ground capacitances, $\mathcal{L}_\textrm{R}$ from the readout cavity and $\mathcal{L}_\textrm{c}$ due to the coupling capacitances $C_\textrm{c}$. The node flux variables at various points in the circuit are denoted by $\varphi_i=2\pi\Phi_i/\Phi_0$. We mark the phase difference across the junctions in the array, $\theta_{i}=\varphi_i-\varphi_{i-1}$, and across the phase-slip junction, $\theta_0=\varphi_\textrm{N}-\varphi_{0}$, in Fig.~\ref{fig:circuit_choice}(a). Similarly, the voltage drop across each junction is given by the phase evolution equation $\dot{\theta}_i=\dot{\varphi}_i-\dot{\varphi}_{i-1}=2\pi V_i/\Phi_0$ where $\Phi_0=h/2e$ the superconducting flux quantum. We will use subscripts $\textrm{j, p}$ for JJA and the phase-slip junction coordinates, respectively. For the Lagrangian of the symmetric circuit~\cite{viola2015collective,ferguson2013symmetries} (setting $\hbar=1$), we have, 
\begin{align}
    \mathcal{L}&=\mathcal{L}_\textrm{PS}+\mathcal{L}_{\textrm{JJA}}+\mathcal{L}_{\textrm{g}}+\mathcal{L}_{\textrm{R}}+\mathcal{L}_{\textrm{c}},
\end{align}
where, ignoring the harmonic oscillator Lagrangian $\mathcal{L}_\textrm{R}$ due to triviality, we have
\allowdisplaybreaks{
\begin{align}
    \mathcal{L}_\textrm{PS}&=\frac{1}{16E_{\textrm{C}}}(\dot{\varphi_\textrm{N}}-\dot{\varphi_0})^2-E_{\textrm{J}_\textrm{p}}\cos\big(\sum_{i=1}^N\theta_i+\varphi_\mathrm{ext}\big)\\
    \mathcal{L}_{\textrm{JJA}}&=\sum_{i=1}^N\frac{1}{16E_{\textrm{C}_\textrm{j}}}(\dot{\varphi}_i-\dot{\varphi}_{i-1})^2-E_{\textrm{J}_\textrm{j}}\cos(\theta_i)\\
    \mathcal{L}_{\textrm{g}}&=\sum_{i=1}^{N-1} \frac{\dot{\varphi_i}^2}{16E_{\textrm{C}_\textrm{g,j}}}+\sum_{i=0,N} \frac{\dot{\varphi_i}^2}{16E_{\textrm{C}_\textrm{g,p}}}\\
    \mathcal{L}_{\textrm{c}}&=\frac{1}{16E_{\textrm{C}_\textrm{c}}}(\dot{\varphi}_{-1}-\dot{\varphi_0})^2+\frac{1}{16E_{\textrm{C}_\textrm{c}}}(\dot{\varphi}_{-2}-\dot{\varphi}_\textrm{N})^2\nonumber\\
    &=\frac{\dot{\varphi}^2_0}{16E_{\textrm{C}_\textrm{c}}}+\frac{\dot{\varphi}^2_{-1}}{16E_{\textrm{C}_\textrm{c}}}-\frac{\dot{\varphi}_0\dot{\varphi}_{-1}}{8E_{\textrm{C}_\textrm{c}}}\nonumber\\&\quad+\frac{\dot{\varphi}^2_\textrm{N}}{16E_{\textrm{C}_\textrm{c}}}+\frac{\dot{\varphi}^2_{-2}}{16E_{\textrm{C}_\textrm{c}}}-\frac{\dot{\varphi}_\textrm{N}\dot{\varphi}_{-2}}{8E_{\textrm{C}_\textrm{c}}}.\label{eq:float-float}
\end{align}
}
The definitions and values of capacitive energies used here are given in Table~\ref{tab:circuit_params}. The terms associated with $\dot{\varphi}_i (\theta_i)$ are the kinetic (potential) energy terms. Note that for the asymmetric circuit in Fig.~\ref{fig:circuit_choice}(b), we can impose $\dot{\varphi}_{-2}=0$ and $C_\textrm{c}=0$ on one end of the circuit. First, we expand on the derivation for the symmetric case in Refs.~\cite{ferguson2013symmetries,viola2015collective} for completeness, and then we impose these conditions to derive the Lagrangian for the asymmetric case.

We use the definition of phase difference quadratures and the fluxoid quantization, 
\begin{align}
\sum_{l=1}^m\theta_l&=\varphi_m-\varphi_0\\
\sum_{m=0}^N \theta_m+\varphi_\mathrm{ext}&=2\pi z, \ z\in\mathbb{Z},
\end{align}
 with an external flux choice of $\varphi_{\mathrm{ext}}=\pi$ as per the main text. Writing the the Lagrangian in this new basis follows $\partial \mathcal{L}/\partial \dot{\varphi}_0=0$, yielding the expression,
\begin{align}
\dot{\varphi}_0&=E_{\textrm{t}}\Big(\frac{\dot{\varphi}_{-1}}{E_{\textrm{C}_\textrm{c}}}+\frac{\dot{\varphi}_{-2}}{E_{\textrm{C}_\textrm{c}}}-\sum_{l=1}^{N}\frac{\dot{\theta}_l}{E_{\textrm{C}_\textrm{c}}}-\sum_{l=1}^{N}\frac{\dot{\theta}_l}{E_{\textrm{C}_\textrm{g,p}}}\nonumber\\&\quad-\sum_{i=1}^{N-1}(N-i)\frac{\dot{\theta}_i}{E_{\textrm{C}_\textrm{g,j}}}\Big)
\end{align}
where
\begin{align}
E_\textrm{t}=\Big(\frac{2}{E_{\textrm{C}_\textrm{c}}}+\frac{N-1}{E_{\textrm{C}_\textrm{g,j}}}+\frac{2}{E_{\textrm{C}_\textrm{g,p}}}\Big)^{-1}\label{eq:tot_cap}
\end{align}
is the total capacitive energy of the circuit due to the parasitic ground capacitances and coupling capacitances. Using the expression for $\dot{\varphi}_0$, we get $\mathcal{L}_\textrm{c}+\mathcal{L}_\textrm{g}=$,
\begin{align}
&\frac{\big(\dot{\varphi}^2_{-1}+\dot{\varphi}^2_{-2}\big)}{16E_{\textrm{C}_\textrm{c}}}\Big(1-\frac{E_\textrm{t}}{E_{\textrm{C}_\textrm{c}}}\Big)+\frac{\dot{\varphi}_{-1}\dot{\varphi}_{-2}}{8E_{\textrm{C}_\textrm{c}}}\Big(\frac{E_\textrm{t}}{E_{\textrm{C}_\textrm{c}}}\Big)\nonumber\\
&+\sum_{i=1}^N\Big(\frac{\dot{\varphi}_{-1}\dot{\theta}_i}{E_{\textrm{C}_\textrm{c}}}+\frac{\dot{\varphi}_{-2}\dot{\theta}_i}{E_{\textrm{C}_\textrm{c}}}\Big)\Big(\frac{E_\textrm{t}}{8E_{\textrm{C}_\textrm{c}}}+\frac{E_\textrm{t}}{8E_{\textrm{C}_\textrm{g,p}}}\Big)\nonumber\\&+\sum_{i=1}^{N-1}\Big(\frac{\dot{\varphi}_{-1}\dot{\theta}_i}{E_{\textrm{C}_\textrm{c}}}+\frac{\dot{\varphi}_{-2}\dot{\theta}_i}{E_{\textrm{C}_\textrm{c}}}\Big)\frac{(N-i)E_\textrm{t}}{8E_{\textrm{C}_\textrm{g,j}}}\nonumber\\
&-\sum_{i=1}^N\frac{\dot{\varphi}_{-2}\dot{\theta}_i}{8E_{\textrm{C}_\textrm{c}}}+\Big(\sum_{i=1}^N\dot{\theta}_i\Big)^2\Bigg(\frac{1}{16E_{\textrm{C}_\textrm{c}}}+\frac{1}{16E_{\textrm{C}_\textrm{g,p}}}\Bigg)\nonumber\\&\quad\times\Bigg(1-\frac{E_\textrm{t}}{E_{\textrm{C}_\textrm{c}}}-\frac{E_\textrm{t}}{E_{\textrm{C}_\textrm{g,p}}}\Bigg)+\frac{\sum_{i=1}^{N-1}\big(\sum_{l=1}^i\dot{\theta}_l\big)^2}{16E_{\textrm{C}_\textrm{g,j}}}\nonumber\\
&-2\sum_{l=1}^N \sum_{l'=1}^{N-1}\dot{\theta}_l\dot{\theta}_{l'}(N-l')\Bigg(\frac{E_\textrm{t}}{8E_{\textrm{C}_\textrm{c}}E_{\textrm{C}_\textrm{g,j}}}+\frac{E_\textrm{t}}{8E_{\textrm{C}_\textrm{g,p}}E_{\textrm{C}_\textrm{g,j}}}\Bigg)\nonumber\\
&-\frac{E_\textrm{t}}{16}\Bigg[\frac{\sum_{i=1}^{N-1}(N-i)\theta_i}{E_{\textrm{C}_\textrm{g,j}}}\Bigg]^2.
\end{align}
We will eventually use the scalar voltage values for $\dot{\varphi}_{-1},\dot{\varphi}_{-2}$, and hence we can ignore any term solely dependent on these variables. The terms independent of $\dot{\varphi}_{-1},\dot{\varphi}_{-2}$ are combined as $\sum_{l=1}^{N}\sum_{l=1}^{N}\mathcal{G}_{ll'}\dot{\theta}_l\dot{\theta}_{l'}$ where,
\begin{align}
\mathcal{G}_{ll'}&=\Big[\frac{\textrm{N}-\text{max}\{l,l'\}}{16E_{\textrm{C}_\textrm{g,j}}}+\frac{1}{16E_{\textrm{C}_\textrm{c}}}+\frac{1}{16E_{\textrm{C}_\textrm{g,p}}}\Big]\nonumber\\&\quad\times\Big[1-\frac{E_\textrm{t}}{E_{\textrm{C}_\textrm{c}}}-\frac{E_\textrm{t}}{E_{\textrm{C}_\textrm{g,p}}}-(\textrm{N}-\text{min}\{l,l'\})\frac{E_\textrm{t}}{E_{\textrm{C}_\textrm{g,j}}}\Big]
\end{align}
This final expression matches Refs.~\cite{ferguson2013symmetries,viola2015collective} for an ordered array. 
\subsection{Collective modes.} 
Now, we define the collective modes for the fluxonium circuit, $\{\phi,\xi_1,...,\xi_{\textrm{N}-1}\}$ such that 
\begin{align}
    \theta_m=\phi/N+\sum_\mu W_{\mu m}\xi_\mu,
\end{align}
and inversely,
\begin{align}
    \phi&=\sum_{m=1}^N\theta_m,\quad \xi_\mu=\sum_{m=1}^N W_{\mu m}\theta_m.
\end{align}
Here, $\phi$ is called the superinductance mode or the \emph{qubit} mode, while $\xi_\mu$ denotes the parasitic modes indexed by $\mu\in\{1,..,N-1\}$~\cite{ferguson2013symmetries}. The matrix $W$ is semi-orthogonal, with dimensions $(N-1)\times N$, and is given by $\sum_m W_{\mu m}W_{\nu m}=\delta_{\mu \nu}$. Its row sum is zero, $\sum_mW_{\mu m}=0$. Thus, the following choice
\begin{align}
    W_{\mu m}=\sqrt{\frac{2}{N}}\cos{\frac{\pi\mu(m-1/2)}{N}},
\end{align}
is observed in~\cite{ferguson2013symmetries} to derive the Lagrangian for the symmetric circuit. The choice of these new variables highlights the collective modes describing the low-energy physics as illustrated in~\cite{catelani2011relaxation,koch2009charging,manucharyan2009fluxonium}. We now split the combined Lagrangian $\mathcal{L}=\mathcal{T}-\mathcal{U}$ into kinetic energy $\mathcal{T}$ and potential energy $\mathcal{U}$ terms in the basis of collective modes as,
\begin{align}
\mathcal{T}=&\Big[\Big(\frac{\dot{\varphi}_{-1}}{E_{\textrm{C}_\textrm{c}}}+\frac{\dot{\varphi}_{-2}}{E_{\textrm{C}_\textrm{c}}}\Big)\Big(\frac{E_{\textrm{t}}}{8E_{\textrm{C}_\textrm{c}}}+\frac{E_{\textrm{t}}}{8E_{\textrm{C}_\textrm{g,p}}}\Big)-\frac{\dot{\varphi}_{-2}}{8E_{\textrm{C}_\textrm{c}}}\Big]\dot{\phi}\nonumber\\
&+\sum_{l=1}^{N-1}(N-l)\frac{E_{\textrm{t}}}{8E_{\textrm{C}_\textrm{g,j}}}\Big(\frac{\dot{\varphi}_{-1}}{E_{\textrm{C}_\textrm{c}}}+\frac{\dot{\varphi}_{-2}}{E_{\textrm{C}_\textrm{c}}}\Big)\Big](\dot{\phi}/N\nonumber\\&+\sum_\mu W_{\mu l}\dot{\xi}_\mu)
  +\sum_{l=1}^N\sum_{l'=1}^N(\dot{\phi}/N+\sum_\mu W_{\mu l}\dot{\xi}_\mu)(\dot{\phi}/N\nonumber\\
&+\sum_\mu W_{\mu l'}\dot{\xi}_\mu)\times\Big[\frac{\textrm{N}-\text{max}\{l,l'\}}{16E_{\textrm{C}_\textrm{g,j}}}+\frac{1}{16E_{\textrm{C}_\textrm{c}}}\nonumber\\&+\frac{1}{16E_{\textrm{C}_\textrm{g,p}}}\Big]\times\Big[1-\frac{E_\textrm{t}}{E_{\textrm{C}_\textrm{c}}}-\frac{E_\textrm{t}}{E_{\textrm{C}_\textrm{g,p}}}-(\textrm{N}\nonumber\\&-\text{min}\{l,l'\})\frac{E_\textrm{t}}{E_{\textrm{C}_\textrm{g,j}}}\Big]\label{eq:kin-energy}\\
    \mathcal{U}&=-E_{\textrm{J}_\textrm{p}}\cos(\phi)-\sum_{l=1}^NE_{\textrm{J}_\textrm{j}}\cos\Big(\phi/N+\sum_\mu W_{\mu l}\xi_\mu\Big)\label{eq:pot-energy}
\end{align}
\subsection{Circuit modifications and assumptions}
From here on, we define a sum over $m,n$ as running from $1$ to $N$, while the sum over $\mu,\nu$ runs from $1$ to $N-1$. To simplify the Lagrangian for the different circuits, we recall that $\sum_m W_{\mu m}=0$ and the semi-orthogonal matrix condition $\sum_m W_{\mu m}W_{\nu m}=\delta_{\mu\nu}$, yielding the identities,
\begin{align}
 \sum_{n=1}^m W_{\mu n}&=-\sum_{n=m+1}^N W_{\mu n}\\
 \sum_{m=1}^N mW_{\mu m}&=-\frac{c_\mu o_{\mu}}{\sqrt{2N}s_\mu^2}\\
 \sum_{m=1}^N m^2W_{\mu m}&=\frac{c_\mu}{\sqrt{2N}s_\mu^2}[N(-1)^\mu-o_\mu]
\end{align}
where $o_\mu=\frac{[1-(-1)^\mu]}{2}, c_\mu=\cos{\frac{\pi\mu}{2N}}, s_\mu=\sin{\frac{\pi\mu}{2N}}$. We now simplify the kinetic energy and potential energy terms from Eqs.~\ref{eq:kin-energy}-\ref{eq:pot-energy} for the different circuits using these identities.
\paragraph{Linear approximation}
We discuss the simplification under the linear assumption on the parasitic modes $\mu$. This includes only linear terms from the Taylor expansion of the cosine ($\cos{x}\sim 1-\frac{x^2}{2}$) in Eq.~\ref{eq:pot-energy}, yielding (up to a constant term)
\allowdisplaybreaks{
\begin{align}
\mathcal{U}&=E_{\textrm{J}_\textrm{p}}\cos(\phi)+\frac{E_{\textrm{J}_\textrm{j}}}{2N}\phi^2+\frac{E_{\textrm{J}_\textrm{j}}}{2}\sum_{\mu}\xi_\mu^2
    \end{align}
}

The potential energies of both circuits remain the same $\mathcal{U}=\mathcal{U}_\textrm{sym} \textrm{(symmetric circuit)}=\mathcal{U}_\textrm{asym}$ \textrm{(asymmetric circuit)}.
\paragraph{Symmetric circuit.} For the circuit in Fig.~\ref{fig:circuit_choice}(a), let the Lagrangian be denoted by $\mathcal{L}_\textrm{sym}=\mathcal{T}_\textrm{sym}-\mathcal{U}_\textrm{sym}$ and impose the conditions $\dot{\varphi}_{-1}=2\textrm{eV}, \dot{\varphi}_{-2}=-2\textrm{eV}$.
\begin{align}
\mathcal{T}_\textrm{sym}&=\frac{\dot{\phi}\textrm{eV}}{4E_{\textrm{C}_\textrm{c}}}+\Big[(M_{00}+G_{00})\dot{\phi}^2+2\sum_{\mu}(M_{0\mu}\nonumber\\&\quad+G_{0\mu})\dot{\phi}\dot{\xi_\mu}+\sum_{\mu,\nu}(M_{\mu\nu}+G_{\mu\nu})\dot{\xi_\mu}\dot{\xi_\nu}\Big].\end{align}
Here, $M$ comes from the phase-slip junction and JJA, while $G$ comes from the coupling and ground capacitances. These coefficients are given by:
\begin{align}
M_{00}&=\frac{1}{E_{C_\textrm{tot}}}+\frac{1}{NE_{\textrm{C}_\textrm{j}}}, M_{0\mu}=0,    M_{\mu\nu}=\frac{\delta_{\mu\nu}}{E_{\textrm{C}_\textrm{j}}}\\
G_{00}&=\frac{1}{64E_{\textrm{t}}}\Big[1-\frac{2}{3}\frac{(N^2-1)E_\textrm{t}}{NE_{\textrm{C}_\textrm{g,j}}}\Big]\\
G_{0\mu}&=-\frac{c_\mu o_{\mu+1}}{32E_{\textrm{C}_\textrm{g,j}}\sqrt{2N}s_\mu^2}\\
G_{\mu\nu}&=\frac{1}{64E_{\textrm{C}_\textrm{g,j}}s_\mu^2}\Big[\delta_{\mu\nu}-\frac{E_{\textrm{t}}}{E_{\textrm{C}_\textrm{g,j}}}\frac{2c_\mu c_\nu o_\mu o_\nu}{N s_\nu^2}\Big].
\end{align}
Note that $C_\textrm{tot}=\textrm{C}_\textrm{p}+C$ is the total capacitance used for the qubit mode. The quantity $G_{0\mu}$ decides the qubit-parasitic coupling strength ($g_{\phi\mu}$) in the final Hamiltonian, while $G_{00}, M_{00}, M_{\nu\nu},  G_{\nu\nu}$ contribute to the charging energies of the qubit and parasitic modes~\cite{viola2015collective}, discussed in Sec.~\ref{sec:expressions} and App.~\ref{app:coupling}. We can see that only the parasitic modes with even index $\mu$ interact with the qubit since $G_{0\mu}=0$ for $\mu\in 2\mathbb{Z}+1$. The Hamiltonian for the symmetric case, thus, will not couple the odd modes (especially, parasitic mode $\mu=1$) with the qubit~\cite{viola2015collective}.

\paragraph{Asymmetric circuit.} For the circuit in Fig.~\ref{fig:circuit_choice}(b), let the Lagrangian be denoted by $\mathcal{L}_\textrm{asym}=\mathcal{T}_\textrm{asym}-\mathcal{U}_\textrm{asym}$ and impose the conditions $\dot{\varphi}_{-1}=4\textrm{eV}, \dot{\varphi}_{-2}=0$. In addition, $C_\textrm{c}=0$ on one end. The changes in the derivation include:
\begin{align}
\dot{\varphi}_0&=E_{\textrm{t}}\Big(\frac{\dot{\varphi}_{-1}}{E_{\textrm{C}_\textrm{c}}}-\sum_{l=1}^{N}\frac{\dot{\theta}_l}{E_{\textrm{C}_\textrm{g,p}}}-\sum_{i=1}^{N-1}(N-i)\frac{\dot{\theta}_i}{E_{\textrm{C}_\textrm{g,j}}}\Big)\\
E_\textrm{t}&=\Big(\frac{1}{E_{\textrm{C}_\textrm{c}}}+\frac{N-1}{E_{\textrm{C}_\textrm{g,j}}}+\frac{2}{E_{\textrm{C}_\textrm{g,p}}}\Big)^{-1}\\
\mathcal{L}_\textrm{c}&+\mathcal{L}_\textrm{g}
=\nonumber\\&\sum_{i=1}^N\frac{\dot{\varphi}_{-1}\dot{\theta}_i}{E_{\textrm{C}_\textrm{c}}}\Big(\frac{E_\textrm{t}}{8E_{\textrm{C}_\textrm{g,p}}}\Big)+\sum_{i=1}^{N-1}\frac{\dot{\varphi}_{-1}\dot{\theta}_i}{E_{\textrm{C}_\textrm{c}}}\frac{(N-i)E_\textrm{t}}{8E_{\textrm{C}_\textrm{g,j}}}\nonumber\\
&+\sum_{l=1}^{N}\sum_{l=1}^{N}\mathcal{G}_{ll'}\dot{\theta}_l\dot{\theta}_{l'},\\
\mathcal{G}_{ll'}&=\Big[\frac{\textrm{N}-\text{max}\{l,l'\}}{16E_{\textrm{C}_\textrm{g,j}}}+\frac{1}{16E_{\textrm{C}_\textrm{g,p}}}\Big]\nonumber\\&\quad\times\Big[1-\frac{E_\textrm{t}}{E_{\textrm{C}_\textrm{g,p}}}-(\textrm{N}-\text{min}\{l,l'\})\frac{E_\textrm{t}}{E_{\textrm{C}_\textrm{g,j}}}\Big]\\
\mathcal{T}_\textrm{asym}&=\frac{\dot{\phi}\textrm{eV}}{4E_{\textrm{C}_\textrm{c}}}\Big[1-\frac{E_\textrm{t}}{E_{\textrm{C}_\textrm{c}}}\Big]+\frac{E_{\textrm{t}}}{2E_{\textrm{C}_\textrm{g,j}}E_{\textrm{C}_\textrm{c}}} \sum_\mu\frac{c_\mu o_\mu}{\sqrt{2N}s_\mu^2} \dot{\xi_\mu}\textrm{eV}\nonumber\\&+(M_{00}+G_{00}')\dot{\phi}^2-2(M_{0\mu}+G_{0\mu}')\dot{\phi}\dot{\xi_\mu}\nonumber\\&+\sum_{\mu,\nu}(M_{\mu\nu}+G_{\mu\nu})\dot{\xi_\mu}\dot{\xi_\nu}\\
G_{00}'&=\frac{1}{64E_{\textrm{t}}}\Bigg[1-\frac{2}{3}\frac{(N^2-1)E_\textrm{t}}{NE_{\textrm{C}_\textrm{g,j}}}-\frac{E_\textrm{t}}{E_{\textrm{C}_\textrm{c}}^2}\Bigg],\\
G_{0\mu}'&=-\frac{c_\mu}{32E_{\textrm{C}_\textrm{g,j}}\sqrt{2N}s_\mu^2}\Bigg[o_{\mu+1}-\frac{E_\textrm{t}}{E_{\textrm{C}_\textrm{c}}}o_{\mu}\Bigg].
\end{align}
We can see that the parasitic modes with any index $\mu$ interact with the qubit as $G_{0\mu}'\neq 0\quad \forall\quad\mu$. The Hamiltonian for the asymmetric case, thus, will couple to all parasitic modes, even and odd (especially parasitic mode $\mu=1$), with the qubit.

The derived Lagrangians provide the Hamiltonians for each circuit. Ref.~\cite{viola2015collective} demonstrated that the symmetric circuit's Lagrangian, \( \mathcal{L}_\textrm{sym} \), under certain assumptions on \( N \), yields the Hamiltonian in Eqs.~\ref{Hamiltonian_total}-\ref{eq:int_hamiltonian}, where the coupling between the qubit and odd modes is zero. In contrast, we claim the asymmetric circuit's Lagrangian, \( \mathcal{L}_\textrm{asym} \), results in a Hamiltonian with non-zero couplings between the qubit and all parasitic modes (odd and even). Thus, the symmetric case eliminates coupling to certain parasitic modes, including the lowest-frequency mode (\( \mu = 1 \)), while the asymmetric case lacks this advantage. Given the parasitic frequency spectrum in Fig.~\ref{fig:meas_circuit}(c) and the discussion in Sec.~\ref{sec:expressions}, the symmetric circuit's Floquet landscape is expected to exhibit fewer parasitic MISTs than the asymmetric circuit. In this work, we are interested in identifying whether or not PMIST can be detrimental, even under the most conservative circuits where it may be avoided. Therefore, we focus exclusively on the symmetric circuit, as shown in Fig.~\ref{fig:meas_circuit}(c) and Fig.~\ref{fig:circuit_choice}(a) of the main text.

\section{Undriven fluxonium circuit}\label{app:Hamiltonian}
In this appendix, we summarize the expressions for coupling strengths and charging energies used in Eqs.~\ref{Hamiltonian_total}-\ref{eq:int_hamiltonian}. We analyze the effect of variations in circuit parameters on these quantities. This analysis is used in Sec.~\ref{sec:expressions} to comment on the dependence of the parasitic effects captured in this work on circuit parameters. We also discuss the quantities related to the corresponding dispersive qubit Hamiltonians and give expressions for the dispersive coupling $\chi_{\phi\mu}$ used in Sec.~\ref{sec:dephasing}.

\subsection{Variation in charging energies and coupling strengths with circuit parameters}\label{app:coupling}
\begin{figure}[t]
    \centering
    \includegraphics[width=\linewidth]{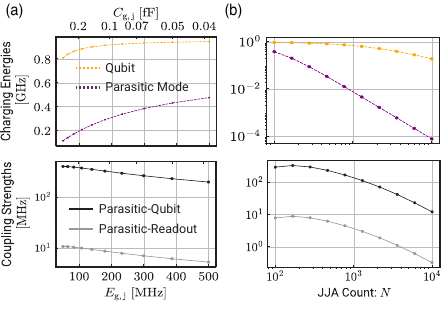}
    \caption{{\bf Dependence of coupling strengths and charging energies on circuit parameters.} (a) parasitic ground capacitance and (b) number of junctions in the array $N$. {\bf (Top row)} The qubit charging energy $E_{\textrm{C}_\textrm{c}}^\phi$ decides the frequency $\omega_{01}$ and the parasitic charging energy $E_{\textrm{c}, \mu}^\textrm{e}$ decides the parasitic mode frequency for mode $\mu=2$. {\bf (Bottom row)} coupling strengths of the parasitic mode to readout ($g_{\mu r}/2\pi$) and qubit ($g_{\phi \mu}/2\pi$), respectively. All plots are obtained under linear JJA approximation.}
    \label{fig:circuit_comp}
\end{figure}

Here, we show the variations of the couplings and charging energies referenced in the main text, using the expressions derived in Ref.~\cite{viola2015collective}, with the number of junctions $N$ and parasitic ground capacitance $C_\textrm{g,j}$, used in Sec.~\ref{sec:expressions}.\begin{figure}[tbh]
    \centering
    \includegraphics[width=\linewidth]{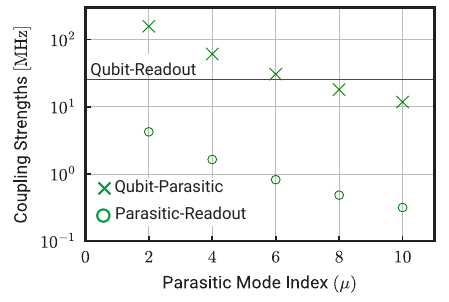}
    \caption{{\bf Absolute values of the coupling strengths.} $g_{\phi \textrm{r}}/2\pi$ (qubit-readout), $g_{\phi\mu}/2\pi$ (qubit-parasitic), $g_{\mu r}/2\pi$ (parasitic-readout), for various circuits. Coupling to odd parasitic modes is zero due to the symmetries of the circuit~\cite{viola2015collective}. The parasitic modes $\mu\in\{2,4,6\}$ couple to the qubit more strongly than the readout.}
    \label{fig:coupling-strength}
\end{figure}
\begin{enumerate}
    \item Qubit charging energy ($\tilde{E}_\textrm{C}^\phi$): The target qubit charging energy is affected by parasitic effects since $\frac{1}{E_\textrm{t}}\neq 0$.\\ $\frac{1}{\bar{E}_\textrm{C}^\phi}=\frac{1}{4E_{\textrm{t}}}\Big(1-\frac{2}{3}\frac{N^2-1}{N}\frac{E_{\textrm{t}}}{E_{\textrm{C}_\textrm{g,j}}}\Big)+\frac{1}{E_{\textrm{C}}}+\frac{1}{NE_{\textrm{C}_\textrm{j}}}$.
    \item Even parasitic-mode charging energy ($E_{\textrm{c},\mu}^\textrm{e}$): The parasitic charging energy is crucial in deciding the parasitic-mode frequency $\omega_{\mu}$ (see Fig.~\ref{fig:meas_circuit}(c)). Because only even parasitic modes (with index $\mu\in 2\mathbb{Z}$) couple to the qubit we will only give expressions for the even modes here.\\ $\frac{1}{E_{\textrm{c},\mu}^{e}}=\frac{1}{E_{\textrm{C}_\textrm{j}}}+\frac{1}{4E_{\textrm{C}_\textrm{g,j}}s_\mu^2}.$ 
    \item Qubit-readout Coupling ($g_{\phi \textrm{r}}$):  This coupling induces all MIST effects.\\ $\frac{g_{\phi \textrm{r}}}{2\pi}=-8E_{C_R}\frac{\tilde{E}_\textrm{C}^\phi}{E_{\textrm{C}_\textrm{c}}} N_{\phi,\mathrm{ZPF}}N_{\mathrm{r,ZPF}}.$
    \item Qubit-parasitic coupling ($g_{\phi\mu}$): This coupling is responsible for PMIST effects, as shown in Fig.~\ref{fig:coupling-Floquet}.\\ $\frac{g_{\phi\mu}}{2\pi}=\frac{4}{\sqrt{2N}} \frac{\tilde{E}^\phi_\textrm{C}E^\textrm{e}_{\textrm{c},\mu}c_\mu}{E_{\textrm{C}_\textrm{g,j}}s_\mu^2} N_{\phi,\mathrm{ZPF}} N_{\mu,\mathrm{ZPF}}.$
    \item Readout-parasitic coupling ($g_{\mu r}$): The parasitic-readout coupling $g_{\mu r}$ only increases the population of the parasitic modes and is directly proportional to $g_{\phi\mu}$. We do not see any significant effects due to the $g_{\mu \textrm{r}}$ quantity in this work. See Table~\ref{tab:params} for the value of $C_\mathrm{R}$.\\ $\frac{g_{\mu r}}{2\pi}=\frac{4}{\sqrt{2N}}\frac{\tilde{E}^\phi_\textrm{C}E^\textrm{e}_{\textrm{c},\mu}c_\mu E_{C_R}}{E_{\textrm{C}_\textrm{g,j}}E_{\textrm{C}_\textrm{c}}s_\mu^2} N_{\mu,\mathrm{ZPF}}N_{\mathrm{r,ZPF}}.$   
    \end{enumerate}
    Fig.~\ref{fig:coupling-strength} shows that the lowest three even modes $\mu=2,4,6$ couple to the qubit more strongly than the readout. This observation is the backbone of our work; we find that because of this relatively large coupling strength, PMIST rates may be significant in fluxonium-based quantum computers. In Fig.~\ref{fig:circuit_comp}, we show the dependence of charging energies and coupling constants on the number of junctions $N$ as well as the ground capacitance $C_\textrm{g,j}$, while keeping all other parameters fixed~\footnote{Note that the assumptions used in Ref.~\cite{viola2015collective} to derive the Hamiltonian were respected when analyzing the variation with $N$. The upper bound on $N\ll 8\pi^2 E_{\mathrm{C}_\mathrm{g,j}}\tilde{E}_\mathrm{C}^\phi\implies N\ll 16000$. Due to high nonlinearity of parasitic modes in the low $N$, and this restriction, our analysis is only valid for the region $10^2<N<10^4$.}. We focus specifically on $g_{\phi\mu}$ and $E_{\textrm{c},\mu}^\textrm{e}$ because the qubit-parasitic coupling is the main mechanism for PMIST, and the parasitic charging energy sets the parasitic frequency.

\paragraph{Variation in coupling strength.}\label{coupling} 
We find that achieving low \( g_{\phi\mu} \) requires a low \( C_\textrm{g,j} \) and a high \( N \) (see Fig.~\ref{fig:circuit_comp}). Increasing \( N \) alters the qubit's target inductance, requiring a proportional increase in \( E_{\textrm{J}_\textrm{j}} \) to maintain the inductive energy (\( E_\textrm{L} = E_{\textrm{J}_\textrm{j}}/N \)). However, \( E_{\textrm{J}_\textrm{j}} \) is limited to a finite value by fabrication constraints, which caps the maximum \( E_\textrm{L} \). This supports the large-\( N \) approximation \( E_{\textrm{c},\mu}^e \approx 4E_{\textrm{C}_\textrm{g,j}}s_\mu^2 \) (Eq.~\ref{eq:parasitic}). Therefore, \( N \) can be optimized to reduce \( g_{\phi\mu} \) while keeping \( E_\textrm{L} \) constant.

However, increasing \( N \) while keeping \( E_\textrm{L} \) and \( E_{\textrm{C}_\textrm{g,j}} \) constant also lowers the parasitic charging energy (see Fig.~\ref{fig:circuit_comp}). For large \( N \) and small \( \mu \), $E_{\textrm{c},\mu}^\textrm{e}$ is
\begin{align}
 E_{\textrm{c},\mu}^\textrm{e} \propto 1/N^2,\label{eq:dep2}
\end{align}
which decreases the parasitic mode frequency \( \omega_\mu \) with increasing \( N \). A low parasitic mode frequency is generally unfavorable for reducing PMIST because it can lead to more allowed PMIST processes, as discussed in the next section. However, if ultimately the coupling \( g_{\phi\mu} \) is negligible, even low-frequency parasitic modes will not contribute to PMIST effects.
\paragraph{Variation in parasitic charging energy.}\label{par-freq} 
A higher parasitic charging frequency relative to the readout frequency reduces the likelihood of PMIST effects (see Sec.~\ref{sec:expressions}). We find that the parasitic charging energies increase with decreasing \( C_\textrm{g,j} \) and \( N \) (see Fig.~\ref{fig:circuit_comp}). From Eqs.~\ref{eq:dep1}-\ref{eq:dep2}, in the large-\( N \), small-\( \mu \) limit, \( \omega_\mu^\textrm{e} \) is inversely proportional to \( N^2 \) and \( C_\textrm{g,j} \), where \( C_\textrm{g,j} \, [\mathrm{fF}] = 19.4 / E_{\textrm{C}_\textrm{g,j}} \, [\mathrm{GHz}] \). Thus, reducing the parasitic ground capacitance \( C_\textrm{g,j} \) increases the charging energy and raises the parasitic mode frequency, which is favorable.

\begin{figure}
    \centering
    \includegraphics[width=\linewidth]{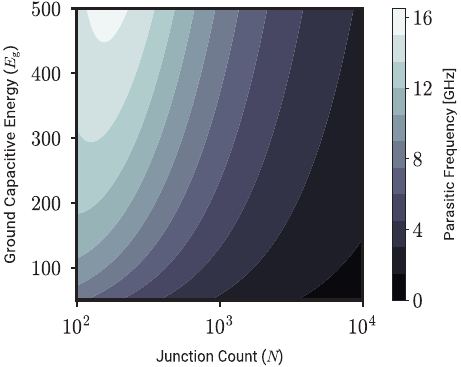}
    \caption{Contour plot of the parasitic frequency $\omega_{\mu=2}$ as a function of circuit parameters, junction count and capacitive energy to ground. For the circuit parameters in the main text, $N=122$ and $E_{\mathrm{C}_\mathrm{g,j}}=194$ GHz, this frequency is equal to $12.06$ GHz. Importantly, for this calculation, unlike Fig.~\ref{fig:circuit_comp}, we increase junction energy of the junctions in the array with $N$ such that the inductive energy $E_\mathrm{L}$ remains constant throughout the plot.}
    \label{fig:par_freq}
\end{figure}

A smaller \( N \) increases \( g_{\phi\mu} \) but also raises the parasitic mode frequencies, widening the gap between \( \omega_\textrm{d} \) and \( \omega_\mu \), which is beneficial. However, decreasing \( N \) also introduces challenges, such as greater nonlinearity of parasitic modes and stronger coupling strengths. Conversely, a higher \( N \) that reduces \( g_{\phi\mu} \) narrows the frequency gap, which is less favorable. 
Parasitic charging energy directly influences parasitic frequency, one of the deciding factors for PMIST events. Thus, in Fig.~\ref{fig:par_freq}, we give a contour plot of this quantity for varying $N$ and $E_{\mathrm{C}_\mathrm{g,j}}$, with an added caveat of constant JJA inductance $E_\mathrm{L}=E_{\mathrm{J}_\mathrm{j}}/N$ as required by fluxonium qubit designs.

These considerations suggest that to minimize PMIST effects (for a given, fixed readout cavity frequency at which PMIST may occur), it is crucial to reduce the parasitic ground capacitance in the JJA. Simultaneously, the junction count \( N \) must be carefully optimized to balance parasitic frequencies and qubit-parasitic couplings, while respecting the assumptions of linearity of parasitic modes.
\subsection{Fluxonium qubit Hamiltonian}
\begin{figure}[t]
    \centering
\includegraphics[width=0.45\textwidth]{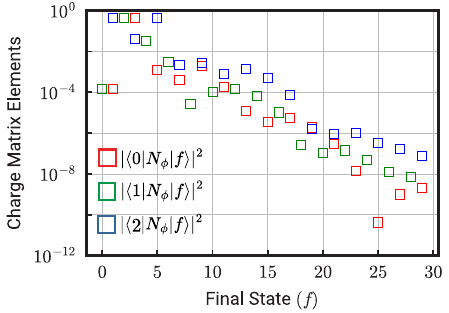}
    \caption{{\bf Charge matrix elements (squared) for the symmetric circuit.} The charge matrix elements between parity-conserving states are zero (points not seen in log plot) due to the symmetry of the cosine potential at $\varphi_\mathrm{ext}=0.5\Phi_0$, where $\Phi_0$ is the flux quantum.}
    \label{charge-matrix}
\end{figure}

We now discuss the parameters related to the fluxonium qubit Hamiltonian through a detailed consideration of its charge matrix elements and the dispersive shifts on the qubit induced by the parasitic modes and readout. The qubit Hamiltonian $\hat H_{\phi}$ (see Eq.~\ref{eq:Hphi}) is diagonalized in the Fock state basis, where we have used the standard bosonic operators.
 \begin{equation} \hat x_\phi=\hat a_\phi+\hat a_\phi^\dagger=\hat N_{\phi}/ N_{\phi,\mathrm{ZPF}}
 \end{equation}
 and 
 \begin{equation} \hat p_\phi=-i(\hat a_\phi-\hat a_\phi^\dagger)=\hat \phi/\phi_{\mathrm{ZPF}}.
 \end{equation}
such that $[\hat x,\hat p]=i$.
\paragraph{Charge matrix elements:}
In Fig.~\ref{charge-matrix}, we plot the charge matrix elements for the qubit mode $\phi$. We observe that with increasing final state ($f$), the charge matrix elements for the ground and first two excited states follow a decreasing trend, approximately exponential. This exponential decrease to $10^{-10}$ motivates our truncation of the fluxonium potential up to $30$ levels (discussed in App.~\ref{app:numerics}) for the Floquet simulations of Sec.~\ref{sec:MIST}.

\begin{figure}[t]
    \centering
    \includegraphics[width=0.45\textwidth]{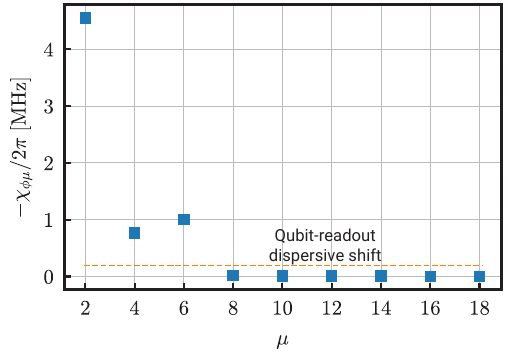}
    \caption{{\bf The dispersive shift ($\chi_{\phi\mu}$) induced on the qubit due to the parasitic mode $\mu$.} The orange reference line represents the dispersive shift ($\sim $) induced by the readout cavity on the qubit ($\chi_{\mu\textrm{r}}$, see Eq.~\ref{eq:qubit-readout-shift}).}
    \label{fig:dispersive-shift}
\end{figure}

\paragraph{Dispersive Hamiltonian}\label{app:dispersive} 
Next, we extract the qubit parameters quoted in Table~\ref{tab:readout_params}, for example, the dispersive shift of the qubit due to the parasitic modes $\chi_{\phi\mu}$ and the readout mode $\chi_{\phi \textrm{r}}$~\cite{viola2015collective}. These variables were used to generate Fig.~\ref{fig:dephasing} in the main text. For this purpose, we first give the qubit Hamiltonian in the dispersive regime~\cite{viola2015collective},
\begin{align}
    H/\hbar&=\omega_\textrm{q}\sigma_\textrm{z}+\sum_{\mu}(\omega_\mu+k_\mu) \hat a_\mu^\dagger \hat a_\mu
    +\omega_\textrm{r} \hat a_\textrm{r}^\dagger \hat a_\textrm{r}\nonumber\\ &\quad +(\chi_{\phi\textrm{r}} \hat a_\textrm{r}^\dagger \hat a_\textrm{r}
    +\sum_{\mu}\chi_{\phi\mu}\hat a_\mu^\dagger \hat a_\mu)\sigma_\textrm{z}\label{eq:dispersive}
\end{align}
where $\omega_\textrm{q}=\omega_{01}$ is the qubit frequency, $\kappa_\mu$ is the Lamb shift on the parasitic mode, while all other variables follow the definitions used in Table~\ref{tab:readout_params}. In the main text, we have used the value of the dispersive shift ($\chi_{\phi\mu}$) due to the parasitic mode in Sec.~\ref{sec:dephasing}, computed using the expression from Ref.~\cite{viola2015collective}. For the parameters in Table~\ref{tab:circuit_params}, we plot $\chi_{\phi\mu}$ for all parasitic modes of the symmetric circuit in Fig.~\ref{fig:dispersive-shift}. We also plot $\chi_{\phi \textrm{r}}$ for reference computed using the following expression (derivation not shown here),
\begin{align}  
\chi_{\phi \textrm{r}}&=16g_{\phi\textrm{r}}^2E_{\textrm{C}_\textrm{R}}^2\sqrt{\frac{E_{\textrm{L}_\textrm{R}}}{32E_{\textrm{C}_\textrm{R}}}}\frac{2\epsilon_{01}}{\epsilon_{01}^2-\omega_\textrm{r}^2}|\langle 0|\hat p_\phi|1 \rangle|^2\nonumber\\
   &+16g_{\phi\textrm{r}}^2E_{\textrm{C}_\textrm{R}}^2\sqrt{\frac{E_{\textrm{L}_\textrm{R}}}{32E_{\textrm{C}_\textrm{R}}}}\Bigg[\sum_l|\langle 0|\hat p_\phi|l \rangle|^2\frac{\epsilon_{0l}}{\epsilon_{0l}^2-\omega_\textrm{r}^2}\nonumber\\&\quad-\sum_l|\langle 1|\hat p_\phi|l \rangle|^2\frac{\epsilon_{1l}}{\epsilon_{1l}^2-\omega_\textrm{r}^2}\Bigg],\label{eq:qubit-readout-shift}
\end{align}
where, $E_{\textrm{C}_\textrm{R}}$ and $E_{\textrm{L}_\textrm{R}}$ are capacitive and inductive energies of the readout resonator, respectively. The dispersive shifts $\chi_{\phi\mu}\ge\chi_{\phi\mathrm{r}}$ for $\mu=\{2,4,6\}$, as expected. 

\section{Driven fluxonium circuit}\label{app:MIST}
Here, we discuss several analysis techniques to study the effects of MIST used in this work. First, we discuss and justify the Hilbert space truncation, as well as approximations used for Floquet simulations in Sec.~\ref{sec:MIST}. Next, we derive the semiclassical Hamiltonian $H_\textrm{s.c.}$ in Eq.~\ref{eq:drive} and discuss other details related to the Floquet simulations.
\subsection{Approximations for numerical modeling}\label{app:numerics}
\begin{figure}[t]
        \centering
        \includegraphics[width=\linewidth]{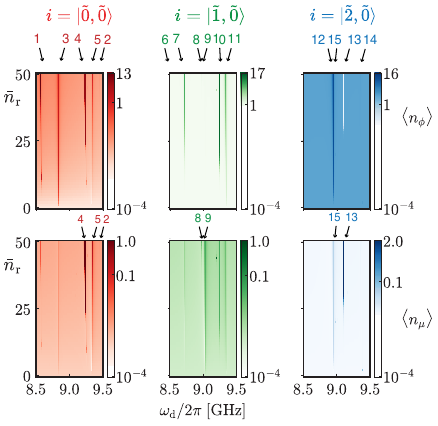}
        \caption{{\bf Truncation.} Floquet landscape of Fig.~\ref{fig:Floquet} computed using a Hilbert space of $30\times 10$ is shown here. The two plots match in the fluxonium and parasitic levels which get excited at the expected frequencies, as quoted in Table.~\ref{tab:PMIST}. These plots are limited to only show transitions up to the first $20$ levels in fluxonium and first $5$ levels in the parasitic mode\footnote{In addition to the transitions shown here, we observe (only) two extra transitions for the $30\times 10$ Floquet simulations, $\ket{\tilde{0},\tilde{0}}\leftrightarrow\ket{\tilde{21},\tilde{0}}$ at $\omega_\textrm{d}=8.60,\bar n_\mathrm{r}=49$ and $\ket{\tilde{2},\tilde{0}}\leftrightarrow\ket{\tilde{20},\tilde{0}}$ at $\omega_\textrm{d}=8.69,\bar n_\mathrm{r}=40$. } for comparison with Fig.~\ref{fig:Floquet} which was obtained using a Hilbert space of $20\times 5$. Same as Fig.~\ref{fig:Floquet}, the figures are plotted in logarithmic scale to pronounce the numbered streaks, the transitions of interest. Only sharp streaks indicate MIST or PMIST while any background change in color can be ignored. For vertical cuts of this plot please see Figs.~\ref{fig:Trans0}-\ref{fig:Trans2}. For a horizontal cut see Fig.~\ref{fig:stark_2D}.
        }
        \label{fig:truncation}
    \end{figure}

We use the following approximations in our work.
\begin{itemize}
    \item \textbf{Restriction to $\mu=2$}. We restrict our analyses to only include the lowest-frequency, even parasitic mode. This mode couples most strongly to the qubit and the readout as evident from Fig.~\ref{fig:coupling-strength}(d). This assumption reduces the Hilbert space size for a feasible study. 

    \item \textbf{Semiclassical readout approximation.} We treat the readout cavity classically as described in Refs.~\cite{dumas2024unified,cohen2023reminiscence,khezri2023measurement}, eliminating the readout mode states from our numerical simulation. This approximation is again necessary to restrict the Hilbert space size to values feasible for numeric study.
    
    \item \textbf{Linear JJA approximation.} We assume that the parasitic modes are linear, due to the large $E_{\textrm{J}_\textrm{j}}/E_{\textrm{C}_\textrm{j}} \sim \mathcal{O}(100)$ ratio. Nonlinear corrections to our results are beyond the scope of this work. For details on how nonlinear corrections affect different circuit energies, we direct the readers to~\cite{viola2015collective}.
    
    \item \textbf{Truncation.} We note that the charge matrix elements connecting the fluxonium qubit ground state to excited states decrease roughly exponentially with increasing excited state number (see Fig.~\ref{charge-matrix}). With this observation and our assumptions above, we truncate the Hilbert space dimensions to $20\times 5$. That is, we assume 20 levels in the fluxonium qubit mode and 5 levels in the parasitic mode. We justify this truncation in Fig.~\ref{fig:truncation} for all Floquet simulations by giving an analogous plot for Fig.~\ref{fig:Floquet} using a numerical simulation in Hilbert space of $30\times 10$.
   \end{itemize}
    Note that for the conclusions drawn in this paper, we are only interested in identifying the existence of PMIST processes. We do not claim to quantify how many such transitions can be present or to enumerate over the entire parameter space. Hence, with this truncation, we only examine the excitations in $0-20$ levels in the fluxonium subspace in Figs.~\ref{fig:Floquet},~\ref{fig:Flo_low},~\ref{fig:coupling-Floquet}. However, in this appendix, we show that the results of truncating the Hilbert space to $20
    \times 5$ levels in the two modes and $30
    \times 10$ match, indicating the convergence of the MIST results. Importantly, both simulations show no higher than an excitation of $\bar n_\mu=2$ in parasitic mode, further motivating our truncation.
\begin{figure}
    \centering
    \includegraphics[width=\linewidth]{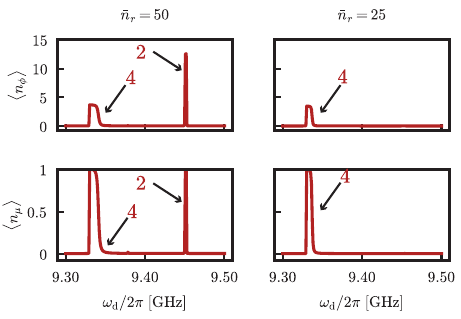}
    \caption{\textbf{Horizontal cut of the Fig~\ref{fig:Floquet}.} This figure highlights the width of transitions 1 (MIST) and 2 (PMIST) along the frequency axis, at $\bar n_r=\{50,25\}$. Notice that at higher drive strengths (i.e, $\bar n_r$) the MIST conditions are satisfied by a larger range of frequencies. For example, the width of the transition $\#4$ increases with increase in $\bar n_r$ while transition $\# 2$ vanishes at $\bar n_r=25$. The intuition behind this behavior is hidden in the fact that the interaction is proportional to the square root of the photon number.}
    \label{fig:stark_2D}
\end{figure}
\subsection{Semiclassical approximation}\label{app:semi-classical}
In this appendix, we derive the semiclassical Hamiltonian $\hat H_\textrm{s.c.}$ (see Eq.~\ref{eq:drive_Ham}) used for the Floquet simulations. The fully-quantum driven readout fluxonium Hamiltonian is given by,
\begin{align}
    \hat H_\textrm{drive}=\hat H \textrm{ (see Eq.~\ref{Hamiltonian_total}) }-i\xi(\hat a_\textrm{r}-\hat a_\textrm{r}^\dagger)\cos{\omega_\textrm{d} t}
\end{align}
where $\omega_\textrm{d}$ is the drive frequency and $\xi\in \mathbb{R}$ is the drive strength. For the semiclassical approximation, we first use the rotating frame transformation $U\hat H_\textrm{drive}U^\dagger-i\dot{U}U^\dagger$ under the unitary $U=e^{-i\hat H_\textrm{r}t}=e^{-i\omega_\textrm{r} \hat a_\textrm{r}^\dagger \hat a_\textrm{r}t}$. This transformation imposes $\hat a_\textrm{r}\rightarrow \hat a_\textrm{r}e^{-i\omega_\textrm{r}t}\equiv\hat{\tilde{a}}$. Now, in the interaction picture, we have $\hat H_\textrm{U} \equiv U\hat H_\textrm{drive}U^\dagger+i\dot{U}U^\dagger$. Dropping terms which oscillate at a rate faster than $\omega_\textrm{d}=\omega_\textrm{r}$),
\begin{align}
   \hat H_\textrm{U} =&\hat H_\phi+\sum_{\mu\in 2\mathbb{Z}}\hat H_\mu+\frac{g_{\phi\mu}\hat N_\phi\hat N_\mu}{N_{\phi,\textrm{ZPF}}N_{\mu,\textrm{ZPF}}}-i\frac{\xi}{2}(\hat a_\textrm{r}+\hat a_\textrm{r}^\dagger)\nonumber\\ 
   &-i\frac{g_{\phi\textrm{r}}\hat N_\phi}{N_{\phi,\textrm{ZPF}}}(\hat {\tilde{a}}_\textrm{r}-\hat {\tilde{a}}_\textrm{r}^\dagger)-i\frac{g_{\mu\textrm{r}}\hat N_\mu}{N_{\mu,\textrm{ZPF}}}(\hat {\tilde{a}}_\textrm{r}-\hat {\tilde{a}}_\textrm{r}^\dagger)
\end{align}
Now we go to the displaced frame $\hat a_\textrm{r}\rightarrow \hat a_\textrm{r}+\alpha$ via transformation under the unitary $U_\alpha=e^{-i\alpha(\hat a_\textrm{r}+\hat a_\textrm{r}^\dagger)}$ such that in the interaction picture we have $\hat H_\textrm{U} \rightarrow U_\alpha\hat H_\textrm{U}U_\alpha^\dagger+i\dot{U}_\alpha U_\alpha^\dagger$: 
\begin{align}
   \hat H_U \rightarrow &\hat H_\phi+\sum_{\mu\in 2\mathbb{Z}}\hat H_\mu+\frac{g_{\phi\mu}\hat N_\phi\hat N_\mu}{N_{\phi,\textrm{ZPF}}N_{\mu,\textrm{ZPF}}}\label{bare}\\ 
   &-i\frac{g_{\phi\textrm{r}}\hat N_\phi}{N_{\phi,\textrm{ZPF}}}(\hat {\tilde{a}}_\textrm{r}-\hat {\tilde{a}}_\textrm{r}^\dagger)-i\frac{g_{\mu\textrm{r}}\hat N_\mu}{N_{\mu,\textrm{ZPF}}}(\hat {\tilde{a}}_\textrm{r}-\hat {\tilde{a}}_\textrm{r}^\dagger)\label{quantum}\\
   &-i\frac{g_{\phi\textrm{r}}\hat N_\phi}{N_{\phi,\textrm{ZPF}}}(\alpha e^{-i\omega_\textrm{r} t}-\alpha^* e^{i\omega_\textrm{r} t})\label{classical1}\\&-i\frac{g_{\mu\textrm{r}}\hat N_\mu}{N_{\mu,\textrm{ZPF}}}(\alpha e^{-i\omega_\textrm{r} t}-\alpha^* e^{i\omega_\textrm{r} t})\label{classical}\\
   &-i\frac{\xi}{2}(\hat a_\textrm{r}+\hat a_\textrm{r}^\dagger)+\dot{\alpha}(\hat a_\textrm{r}+\hat a_\textrm{r}^\dagger)\label{drive}
\end{align}
We choose $\dot{\alpha}=i\frac{\xi}{2} \ \textrm{s.t.} \ \mathrm{Re}(\alpha(t))=0$ to remove the terms in Eq.~\ref{drive}. For the time-dependent simulations emulating the decay of a readout resonator, $\dot\alpha$ is a function of readout decay rate $\kappa_\mathrm{r}$. Under this choice, if $\omega_\mathrm{r}=\omega_\mathrm{d}$ is used, Eqs.~\ref{classical1}-\ref{classical} take the form,
\begin{align}
    \Big[\frac{g_{\phi\textrm{r}}\hat N_\phi}{N_{\phi,\textrm{ZPF}}}+\frac{g_{\mu\textrm{r}}\hat N_\mu}{N_{\mu,\textrm{ZPF}}}\Big]2|\alpha(t)|\cos{\omega_\textrm{d} t}\label{semi-classical}
\end{align}
Now, we go back to the unrotated frame and drop the terms associated with the quantum operators $(\hat{a}_\textrm{r}-\hat{a}_\textrm{r}^\dagger)$ in Eqs.~\ref{quantum} while we replace $|\alpha(t)|$ with the mean value associated with the average number of photons in the readout at any time $\sqrt{\bar n_\textrm{r}(t)}$ in the term~\ref{semi-classical} to be used in $\hat H_\textrm{U}$~\cite{cohen2023reminiscence}, resulting in the final semi-classical form $\hat H_\textrm{s.c.}$.

\subsection{Floquet simulations}
 \begin{figure}[t]
     \centering
     \includegraphics[width=\linewidth]{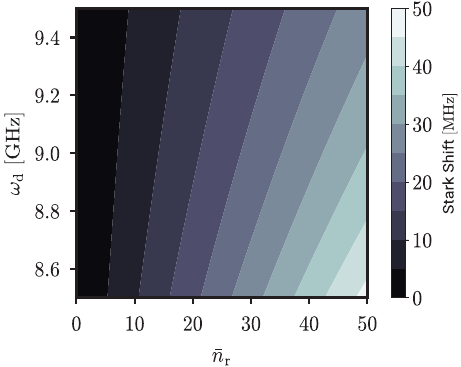}
     \caption{\textbf{Contour plot of the stark shift in the ground state of fluxonium due to readout.} Values are computed using Eq.~\ref{eq:stark} for energy level $i=\ket{\tilde{0},\tilde{0}}$. We have verified that the Stark shift on the excited energy levels $\ket{\tilde{1},\tilde{0}}$ and $\ket{\tilde{2},\tilde{0}}$ are also upper bounded by $50$ MHz for the ranges of $\omega_\textrm{d}$ and $\bar n_\textrm{r}$ considered in this plot.}
     \label{fig:stark-shift}
 \end{figure}
\begin{figure*}
    \centering
    \includegraphics[width=1.0\textwidth]{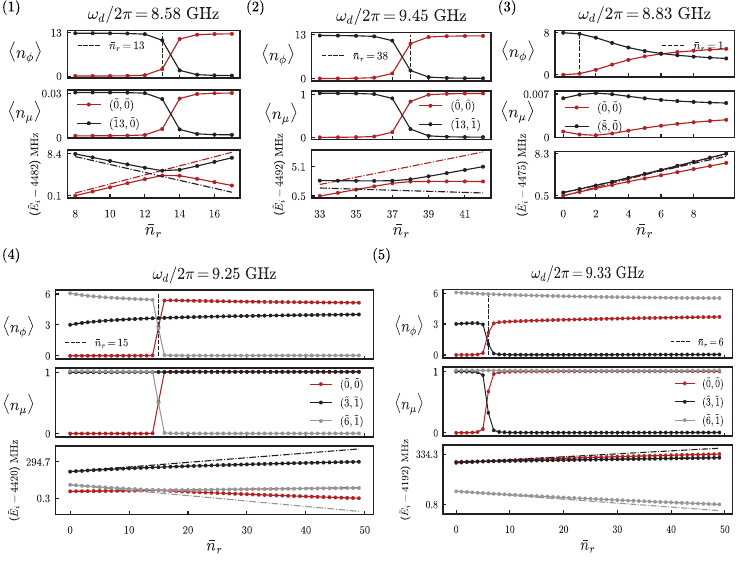}
    \caption{\textbf{MIST processes from Table~\ref{tab:PMIST} involving the $\ket{\tilde{0},\tilde{0}}$ state.} The figure numbering indicates the row index in Table~\ref{tab:PMIST}. (Top row) Fluxonium subspace $\braket{n_\phi}$. (Middle) Parasitic mode subspace $\braket{n_\mu}$. (Bottom) Stark-shifted eigenenergies (dashed) and quasi-energies (solid) from Floquet simulations correspond to the initial states $i$ as per the legend. The y-axis for this plot is in MHz. Floquet results are extracted from numerical data used for Fig.~\ref{fig:Floquet}. Note that MIST in subpanel $(2)$ is split into two figures to capture the two consecutive transitions involved.}
    \label{fig:Trans0}
\end{figure*}
\begin{figure*}
    \centering
    \includegraphics[width=1.0\textwidth]{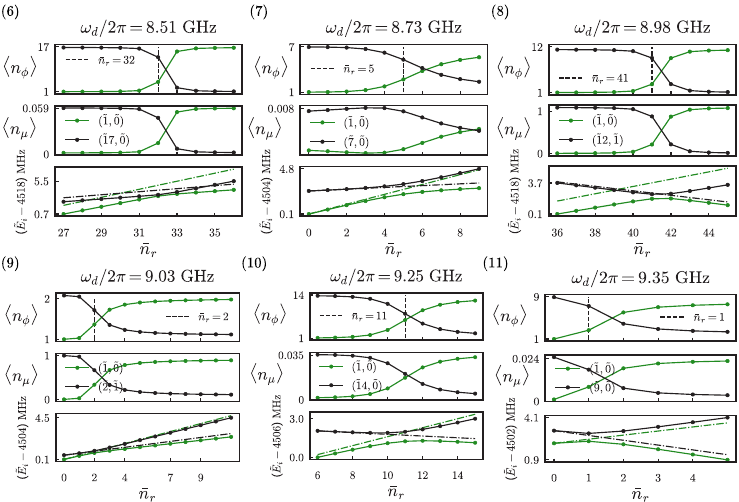}
    \caption{\textbf{MIST processes from Table~\ref{tab:PMIST} involving the $\ket{\tilde{1},\tilde{0}}$ state.} The figure numbering indicates the row index in Table~\ref{tab:PMIST}. (Top row) Fluxonium subspace $\braket{n_\phi}$. (Middle) Parasitic mode subspace $\braket{n_\mu}$. (Bottom) Stark-shifted eigenenergies (dashed) and quasi-energies (solid) from Floquet simulations correspond to the initial state $i$ as per the legend. The y-axis for this plot is in MHz. Floquet results are extracted from numerical data used for Fig.~\ref{fig:Floquet}.}
    \label{fig:Trans1}
\end{figure*}
\begin{figure*}
    \centering
    \includegraphics[width=1.0\textwidth]{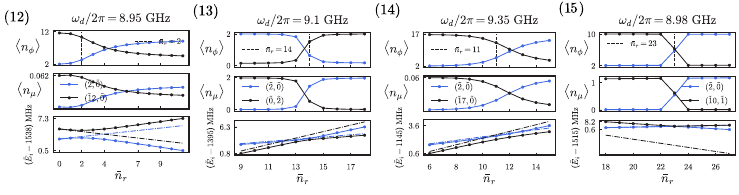}
    \caption{\textbf{MIST processes from Table~\ref{tab:PMIST} involving the $\ket{\tilde{2},\tilde{0}}$ state.} The figure numbering indicates the row index in Table~\ref{tab:PMIST}. (Top row) Fluxonium subspace $\braket{n_\phi}$. (Middle) Parasitic mode subspace $\braket{n_\mu}$. (Bottom) Stark-shifted eigenenergies (dashed) and quasi-energies (solid) from Floquet simulations correspond to the initial state $i$ as per the legend. The y-axis for this plot is in MHz. Floquet results are extracted from numerical data used for Fig.~\ref{fig:Floquet}.}
    \label{fig:Trans2}
\end{figure*}

\paragraph{Stark shift:}\label{app:stark-shift}
In this section, we compute the Stark-shifted eigenenergies that facilitate the prediction of an avoided crossing using a first-order perturbative approach, given $\bar n_\textrm{r}, \omega_\textrm{r}$ in Figs.~\ref{fig:011},~\ref{fig:Trans0}-\ref{fig:Trans2}, and the charge-matrix elements. To observe a state transition, the primary requirements are large coupling, large charge-matrix elements, and small energy differences. The eigenenergies of the states in question are changed with an increase in the number of readout photons or, in this case, the drive strength. Let $\ket{i}$ be a state in the eigenspace of $\hat H_0$ (see Eq.~\ref{eq:bare_ham}). The Stark shift in the energy of state $\ket{i}$ at an average number of readout photons $\bar n_\textrm{r}$ is given by
\begin{align}
    \chi_i(\bar n_\textrm{r})&=2\bar n_\textrm{r}\sum_{f}\tilde{\omega}_{if} \Big[\frac{g_{\phi \textrm{r}}|\braket{i|\hat N_\phi|f}|^2}{N_{\phi,\mathrm{ZPF}}(\omega_\textrm{d}^2-\tilde{\omega}_{if} ^2)}\nonumber\\&\quad\quad+\frac{g_{\mu \textrm{r}}|\braket{i|\hat N_\mu|f}|^2}{N_{\mu,\mathrm{ZPF}}(\omega_\textrm{d}^2-\tilde{\omega}_{if} ^2)}\Big].\label{eq:stark}
\end{align}
Here $\tilde{\omega}_{if}=E_f-E_i$ denotes the energy difference in the eigenenergies of the hyrbidized states $\ket{i}$ and $\ket{f}$. The second term is much smaller than the first term, and hence the qubit-readout coupling strength $g_{\phi \textrm{r}}$ primarily governs this Stark shift.

\paragraph{Population exchange and quasienergies:}\label{app:Floquet-trans}
Below we plot the population exchange and quasi-energy probabilities for all transitions captured in Fig.~\ref{fig:Floquet} and Table~\ref{tab:PMIST} for states $\ket{\tilde{0},\tilde{0}}$ (see Fig.~\ref{fig:Trans0}), $\ket{\tilde{1},\tilde{0}}$ (see Fig.~\ref{fig:Trans1}), and $\ket{\tilde{2},\tilde{0}}$ (see Fig.~\ref{fig:Trans2}). We also use the stark-shifted energies computed from Eq.~\ref{eq:stark} to predict the transitions using a perturbative analysis, plotted in dashed lines with quasienergies. We comment on some special types of transitions explicitly here.
\begin{itemize}
    \item Transitions $(\#3,\#4)$ show a branch bunching scenario, as quoted in Ref.~\cite{dumas2024unified} for the case of positive detuning. Such branch bunching occurs when the drive frequency $(\omega_\textrm{d})$ is equal to the transition frequency ($\omega_{if}$) of two bare fluxonium states $i,f$. In the presence of a parasitic mode, there are \emph{multiple} states ($\ket{\tilde{i},\tilde{\mu}}$ and $\ket{\tilde{f},\tilde{\mu}}$ for various $\mu$) such that $\omega_\textrm{d}=\omega_{if}$ is the resonant frequency for the observation of bunching between the two states.
    \item In transitions $(\#9,\#11,\#12)$ the swapped states do not return to the same population as the initial states. This is because we are plotting the bare fluxonium and parasitic mode operators $\braket{\bar n_\phi},\braket{\bar n_\mu}$, as opposed to the dressed ones (which should show a recovery of the initial state). We make this choice to show the impact of PMIST effects on the bare fluxonium population.
\end{itemize}

\paragraph{Landau-Zener probabilities:}\label{app:LZ}
We compute the Landau-Zener probabilities in Sec.~\ref{sec:LZ} numerically using the quasienergies from the Floquet simulations. In this case, we use a time-dependent readout photon number, where $\bar n_\textrm{r}$ varies as $\bar n_\textrm{r}=50(1-e^{-\kappa t/2})^2$, to emulate change in readout photons from dissipation~\cite{dumas2024unified,khezri2023measurement}. The numerical calculations use the probability for Landau-Zener transitions given in~\cite{ikeda2022floquet}, for an avoided crossing observed between states $\ket{i},\ket{f}$ of
\begin{align}
    P_\textrm{LZ}&=\exp{\Big[-\frac{\pi \Delta_\textrm{ac}^2}{2v}\Big]},\\
    \text{where } v&=\sqrt{2\Delta_\textrm{ac}\Bigg|\frac{d^2\epsilon_f}{d\sqrt{\bar{n}_\textrm{r}(t)}^2}\Bigg|_{t_\textrm{ac}}}\frac{d\sqrt{\bar{n}_\textrm{r}(t)}}{dt}\Bigg|_{t_\textrm{ac}} .
\end{align}
Here, the variable $\epsilon_f$ is the numerically computed quasi-energy obtained from Floquet simulations, while $\Delta_\textrm{ac}$ refers to their quasi-energy difference $\epsilon_{if}=\epsilon_f-\epsilon_i$ at avoided crossing. 

\section{Alternative circuit parameters}\label{Will_circuit}

Our analysis of PMIST so far has focused on systems where the fluxonium qubit-frequency is $\omega_{01} / 2 \pi \sim 30$ MHz. Here, we consider how these processes change when one uses a larger qubit frequency $\omega_{01}/2\pi \sim 300 \ \mathrm{MHz}$, as was recently realized in the experiment of Ref.~\cite{ding_high-fidelity_2023}. For this purpose, we analyze a new set of circuit parameters (see Table~\ref{tab:circuit_params_Will}).

\begin{table}[htb]
\centering
\begin{tabular}{|c|c|c|c|c|c|c|c|c|c|}
    \hline
     $N$ & $\varphi_{\textrm{ext}}$ & $E_{\textrm{J}_\textrm{p}}$ & $E_{\textrm{C}_\textrm{p}}$ & $E_{\textrm{C}}$ & $E_{\textrm{J}_\textrm{j}}$ & $E_{\textrm{C}_\textrm{j}}$ & $E_{\textrm{C}_\textrm{g,j}}$ & $E_{\textrm{C}_\textrm{g,p}}$ & $E_{\textrm{C}_\textrm{c}}$ \\
    \hline
    $102$ & $0.5\Phi_0$ & $6.20$ & $9.74$ & $1.11$ & $81.6$ &$0.74$  & $194$ & $1.94$ & $19.40$ \\
    \hline
\end{tabular}
\caption{\textbf{Circuit parameters inspired by Ref.~\cite{ding_high-fidelity_2023}.} All energies are given in GHz. Here $\Phi_0=h/2e$ denotes the magnetic flux quantum. The capacitive energies $E_{\textrm{C}'}=\frac{19.4}{{C'}(\mathrm{fF})} \ \mathrm{GHz}$ are computed from the corresponding capacitances $C'$.}
\label{tab:circuit_params_Will}
\end{table}

The parasitic mode frequency of the $\mu=2$ mode is $\omega_{\mu=2}/2\pi=15.50 \ \mathrm{GHz}$. The coupling strengths are: $g_{\phi \textrm{r}}/2\pi=37 \ \mathrm{MHz}$, $g_{\phi\mu}/2\pi=432 \ \mathrm{MHz}$, $g_{\mu \textrm{r}}/2\pi=11 \ \mathrm{MHz}$. The plasma frequency is $\omega_{12}/2\pi=5.40 \ \mathrm{GHz}$. 
\begin{figure}[htb]
    \centering
    \includegraphics[width=\linewidth]{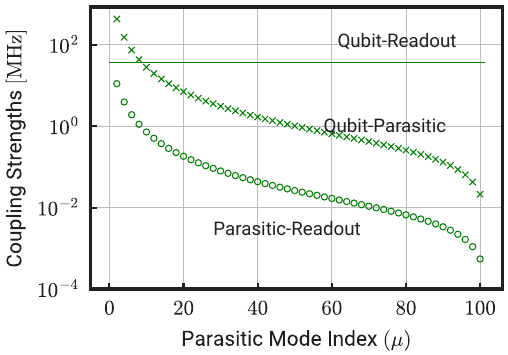}
    \caption{{\bf Magnitudes of the coupling strengths for parameters in Table~\ref{tab:circuit_params_Will} used in the symmetric readout circuit.} $g_{\phi \textrm{r}}/2\pi$ (qubit-readout), $g_{\phi\mu}/2\pi$ (qubit-parasitic), $g_{\mu \textrm{r}}/2\pi$ (parasitic-readout), for various circuits. Coupling to odd parasitic modes is zero due to the symmetries of the circuit. The parasitic modes $\mu\in\{2,4,6\}$ couple to the qubit more or as strongly as the readout.}
    \label{fig:coupling-strength-Will}
\end{figure}
Even though the coupling strengths for these circuit parameters as shown in Fig.~\ref{fig:coupling-strength-Will} are similar to the parameters studied in the main text (see Fig.~\ref{fig:coupling-strength}), since the $\mu=2$ parasitic mode is larger by about $4 \ \mathrm{GHz}$, we expect fewer PMIST processes for the drive frequency range analyzed in Fig.~\ref{fig:Floquet}.

\begin{figure}[htb]
    \centering
    \includegraphics[width=\linewidth]{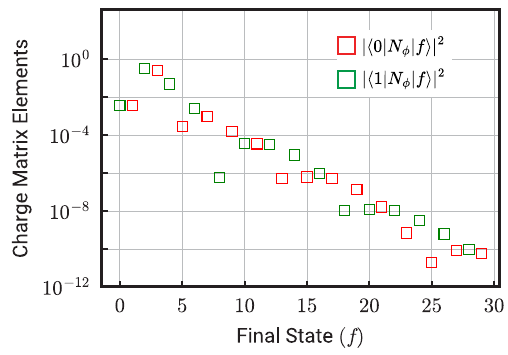}
    \caption{{\bf Charge matrix elements (squared) for parameters in Table~\ref{tab:circuit_params_Will}.} The charge matrix elements between odd-odd or even-even states are zero (points not seen in log plot) due to the symmetry of the cosine potential at $\varphi_\mathrm{ext}=0.5\Phi_0$, where $\Phi_0$ is the flux quantum.}
    \label{fig:charge-matrix-Will}
\end{figure}

\begin{figure*}[htb]
    \centering
    \includegraphics[width=\linewidth]{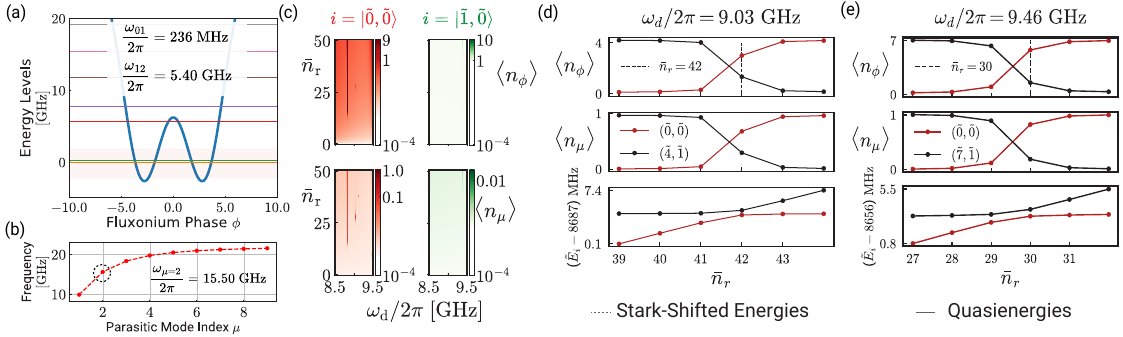}
    \caption{{\bf Floquet analysis with alternate JJA fluxonium parameters inspired by Ref.~\cite{ding_high-fidelity_2023}.} See Table~\ref{tab:circuit_params_Will} for a list of circuit parameters. \textbf{(a)} Fluxonium energy spectrum. Defining $\omega_{ij}$ as the splitting frequency between the fluxonium states $i,j$, we have $\omega_{01}/2\pi=236 \ \mathrm{MHz}, \ \omega_{12}/2\pi=5.40 \ \mathrm{GHz}$. \textbf{(b)} Parasitic mode frequencies. \textbf{(c)} Floquet simulations for the branch analysis of the computational states. The truncation for these simulations uses the same justification and assumptions as App.~\ref{app:numerics}. Same as Fig.~\ref{fig:Floquet}, the Floquet figures are plotted in logarithmic scale to pronounce the numbered streaks, the transitions of interest. Only sharp streaks indicate MIST or PMIST while any background change in color can be ignored. \textbf{(d,e)} Examples of PMIST involving the states $\ket{\tilde{0},\tilde{0}}\leftrightarrow\{\ket{\tilde{4},\tilde{1}},\ket{\tilde{7},\tilde{1}}\}$, with maximum overlap to the corresponding un-hybridized states $\ket{k}_\phi\otimes\ket{n}_{\mu=2}$. \textbf{Top row:} Qubit mode average occupation $\braket{n_\phi}$. \textbf{Middle row:} Parasitic mode average occupation $\braket{n_\mu}$. \textbf{Bottom row:} Quasi-energies (solid) from Floquet simulations showing avoided crossings. Plots are extracted from numerical data used in Fig.~\ref{fig:Floquet}. The data points are connected by lines for visual aid.}
    \label{fig:Floquet1}
\end{figure*}

In addition, Fig.~\ref{fig:charge-matrix-Will} shows that the charge matrix elements of this circuit has a faster decrease with increasing excited state levels. This could be indicative of the fact that such a circuit will see lower MIST effects as observed in Fig.~\ref{fig:Floquet1}(c). This also further supports our assumptions for numerical modeling to be the same as previous Floquet simulations as discussed in App.~\ref{app:numerics}.

Fig.~\ref{fig:Floquet1} shows that indeed, PMIST effects are comparatively less likely for this alternate circuit. The two PMIST processes (out of the four MIST processes) observed in the Floquet profile $\ket{\tilde{0},\tilde{0}}\leftrightarrow\{\ket{\tilde{4},\tilde{1}},\ket{\tilde{7},\tilde{1}}\}$, respectively occur at $\omega_r/2\pi=\{9.03,9.46\}\ \mathrm{GHz},\ \bar n_\textrm{r}=\{42,30\}$, and have the quasi-energy gap of $\Delta_\textrm{ac}=\{0.63,0.60\} \ \mathrm{MHz}$ at the avoided crossings. The explicit transition with quasi-energies is shown in Fig.~\ref{fig:Floquet1}. 

The reduced number of MIST effects in Fig.~\ref{fig:Floquet1} compared to Fig.~\ref{fig:Floquet} can be attributed to the fact that the charge matrix elements 
$|\bra{i}\hat N_\phi\ket{f}|^2$ between computational basis levels $i=0$ (and $i=1$) and the higher fluxonium states ($f$)
decrease faster with increasing $f$ for the new parameter set compared to our original choices (compare Fig.~\ref{fig:charge-matrix-Will} with Fig.~\ref{charge-matrix}).

While the above results are promising, we stress that our analysis here considers a single specific PMIST process (i.e., involving the lowest-frequency array mode with the strongest qubit-parasitic coupling). Other less obvious processes may also play a role, for example, PMIST due to parasitic modes $\mu \geq4$. The Floquet simulation including all relevant parasitic modes is numerically intractable, 
thus motivating the development of other modelling approaches for this problem.
\clearpage
\bibliography{refs.bib} 
\end{document}